# OptAgent: an Agentic AI framework for Intelligent Building Operations

Zixin Jiang[a], Weili Xu[b], Bing Dong[a],*

[a] Department of Mechanical & Aerospace Engineering, Syracuse University, 263 Link Hall, Syracuse, NY 13244, United States
[b] Building Simulation & Design Group, Pacific Northwest National Laboratory, Richland, WA 99354, United States

*Corresponding Author:*
Bing Dong[a],* Email: bidong@syr.edu

**Abstract:**
The urgent need for building decarbonization calls for a fundamental paradigm shift in future autonomous building energy operation, from human-intensive engineering workflows toward intelligence agents that can interact with physics-grounded digital environments. To support this transition, this study proposes an end-to-end agentic AI-enabled Physics-Informed Machine Learning (PIML) environment for scalable building energy modeling, simulation, control and automation. The proposed framework consists of: (1) a modular and physics-consistent PIML digital environment spanning building, Heating, Ventilation, and Air Conditioning (HVAC), distributed energy resources (DER) to support grid-interactive energy management; and (2) an agentic AI layer implemented with 11 specialist agents and 72 Model Context Protocol (MCP) tools to enable end-to-end execution of multi-step energy analytics. A representative case study is presented to demonstrate the execution trace of multi-domain, multi-agent coordination for assessing how system and control upgrades impact energy use, operating cost, thermal comfort, and flexibility. Furthermore, a large-scale benchmark (~4,000 runs) is conducted to systematically evaluate workflow performance in terms of accuracy (planning, agent selection, tool selection, and parameter extraction), token consumption, execution time, and inference cost. The benchmark results quantify the impacts of intelligence mode design, model-size configuration, task complexity, and orchestrator–specialist coordination on overall performance, and provide six key lessons learned for building future agentic AI systems in real-world building energy applications. This work establishes a scalable, physics-grounded foundation for deploying agentic AI in decarbonized and grid-interactive building energy operations, and highlights key research directions toward adaptive self-driving building intelligences.

**Abbreviations**

| | | | |
|---|---|---|---|
| Distributed Energy Resources | DERs | Model Context Protocol | MCP |
| Heating, Ventilation, and Air Conditioning | HVAC | Agent-to-Agent | A2A |
| Artificial Intelligence | AI | Building Information Modeling | BIM |
| Physics-Informed Machine Learning | PIML | Coefficient of Performance | COP |
| Large Language Models | LLMs | Single-Agent Single-Tool | SAST |
| Centralized Single-Stage | C-1 | Single-Agent Multi-Tool | SAMT |
| Centralized Two-Stage | C-2 | Multi-Agent Single-Tool | MAST |
| Decentralized | D | Multi-Agent Multi-Tool | MAMT |
| Large | L | Extra-Large | XL |
| Small | S | Medium | M |

**Symbols**

| | | | |
|---|---|---|---|
| $\lvert \cdot \rvert$ | Cardinality set size | $\subseteq$ | Subset |
| $\cap$ | Intersection | | |

**Notation**

| | | | |
|---|---|---|---|
| $T_{\text{exp}}$ | Expected tool | $T_{act}$ | Actual tool |
| $G_{\text{exp}}$ | Expected agent | $G_{act}$ | Actual agent |
| $S_{\text{exp}}$ | Expected plan trace | $S_{\text{act}}$ | Actual plan trace |
| $K_{exp}$ | Expected tool key | $K_{act}$ | Actual tool key |
| $V_{exp}$ | Expected tool value | $V_{act}$ | Actual tool value |
| $g_i$ | One step agent selection | $\mathcal{T}_i$ | One step tool use |
| $Acc_{tool}$ | Tool accuracy | $Acc_{agent}$ | Agent accuracy |
| $Acc_{plan}$ | Plan accuracy | $Acc_{key}$ | Key accuracy |

## 1. Introduction

Buildings are among the largest energy consumers, accounting for over one-third of total energy use globally[1] and are essential to achieving the 2050 net-zero emissions target. Beyond their energy footprint, buildings also serve as active energy hubs [2], tightly coupled with distributed energy resources, power grid, heating, ventilation, and air conditioning  systems, as well as human mobility and transportation networks, playing a vital role in grid-interactive, multi-energy coordination. Moreover, buildings are the primary indoor environment for human life, where people spend nearly 90% of their time, making them critical for ensuring comfort, health, well-being and resilience.

As a result, modern buildings have evolved into complex ecosystems [3] involving *multi-scale*, *multi-component*, *multi-stakeholder*, and *multi-objective* coordination. This complexity poses significant challenges for modeling, control, diagnostics, analysis and evaluation. Traditionally, these tasks rely heavily on expert knowledge, substantial time investment, and repeated manual effort [4][2]. However, achieving the 2050 net-zero goal requires decarbonizing buildings at a rapid and large scale. For example, it has been estimated that over 10,000 buildings per day would need to be decarbonized in the United States alone over the next 25 years [5]. This urgency calls for a fundamental paradigm shift toward a unified, scalable, and robust framework capable of supporting autonomous decision-making across diverse building applications, thereby reducing human effort and adapting to evolving operating conditions.

To enable such an autonomous framework, we propose a vision built on two foundational pillars, as illustrated in Figure 1:

- **Agentic Artificial Intelligence (AI) as the *brain***, responsible for autonomous, end-to-end decision-making across **muti-agent** workflows; and
- **A Physics-Informed Machine Learning Environment as the *body***, providing a scalable, physics-consistent, and adaptive simulation platform in which AI agents can learn, reasoning, and interact with building energy systems.

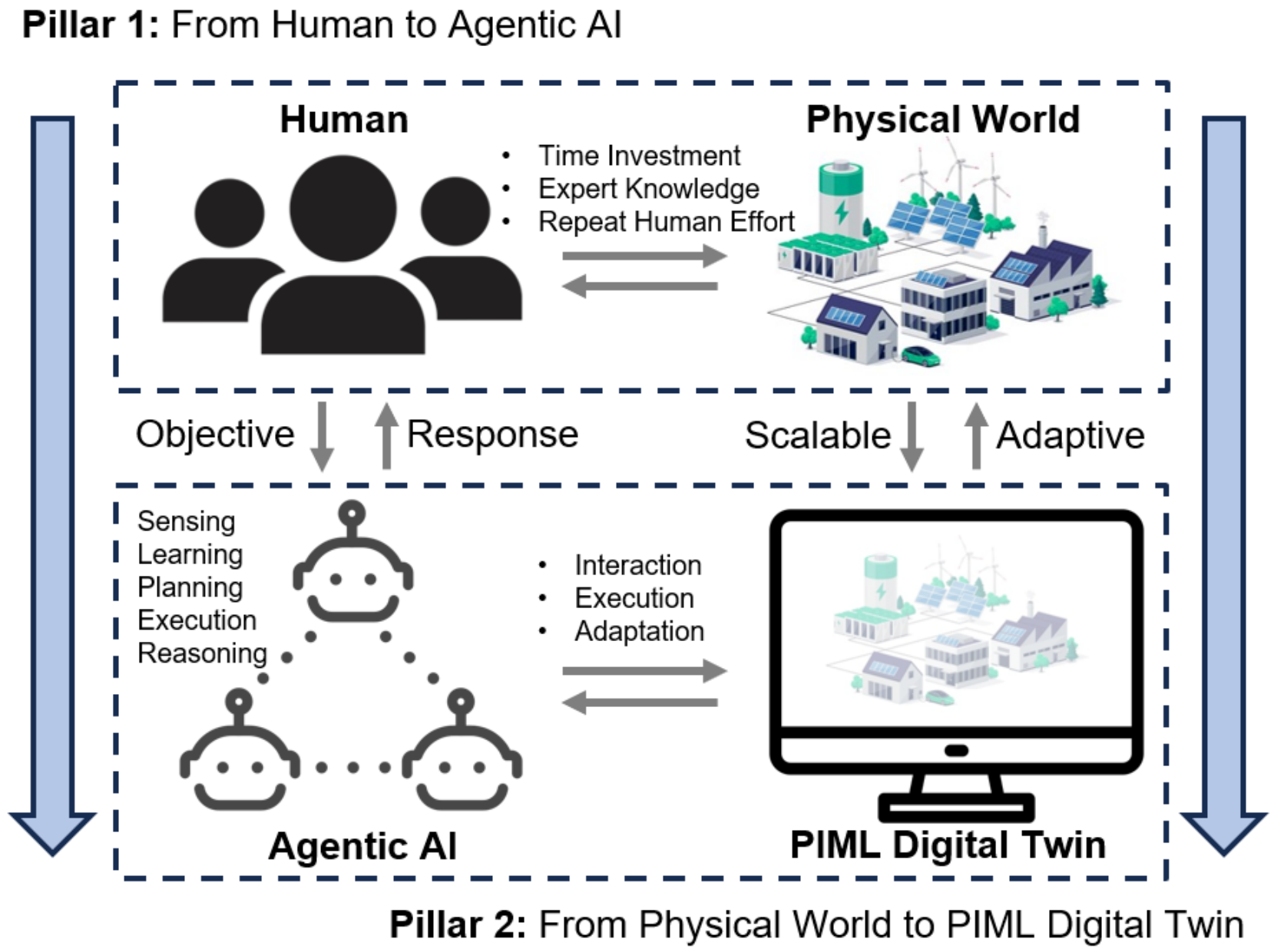

Figure 1. Vision of Agentic AI enabled Physics-informed Machine Learning Framework for Future Autonomous Building Operation. On one hand, agentic AI aims to automate human-intensive tasks through end-to-end decision-

making agents. On the other, a scalable, physics-informed environment enables those agents to learn, reason, and interact with the built energy ecosystem in a physics-consistent and realistic manner.

### 1.1 End-to-End Agentic Artificial Intelligence

Recent advances in large language models (LLMs) have enabled a new class of AI systems, often referred to as **Agentic AI**, which is rapidly gaining industrial adoption due to its potential to reduce human workload, improve scalability, and support autonomous operations. For example, Gartner predicts that by 2028, 33% of enterprise software applications will embed agentic AI, up from less than 1% in 2024 [8]. Reports from Capgemini [9] and Deloitte [10] highlight similar trajectories, and related investments are expected to exceed $749 billion by 2028 [11].

Despite this momentum, current research in intelligent building applications often conflates **Generative AI, AI Agents, and Agentic AI**, leading to conceptual confusion and misaligned system design. To clarify these distinctions, Table 1 compares these paradigms in terms of objectives, interaction patterns, autonomy, memory, and reasoning depth. In summary, agentic AI extends foundation models beyond content generation by introducing planning, memory, tool use, and coordinated decision-making, enabling end-to-end automation of complex workflows.

Table 1 Comparison of Generative AI, AI Agents, and Agentic AI Across Key Capabilities

| **Feature** | **Generative AI** | **AI Agent** | **Agentic AI** |
|---|---|---|---|
| Objective | Content generation based on prompt | Specific task execution using external tools or APIs | Complex workflow and high-level target automation |
| Interaction | User ↔ LLM | User ↔ Agent ↔ Tool | User ↔ Orchestrator ↔ Agents ↔ Environment |
| Flexibility | (Low) Fixed to pretrained abilities | (Medium) Executes defined tasks with limited adaptability | (High) Adapts to new tasks, and coordinates across modules |
| Memory | No memory or short context window | Episodic memory (per task) | Long-term memory, multi-session state |
| Reasoning | Pattern-based reasoning, implicit | Explicit task-level reasoning | Structured reasoning, long-horizon planning, multi-agent coordination |
| Autonomy Level | (Low) Fully user-driven | (Medium) Within a bounded task | (High) Task decomposation and management |
| Tool | None | Tool-enabled | Tool-enabled |

- **Generative AI** refers to systems based on foundation models that generate digital artifacts in response to user prompts [19]. While powerful in content generation, these models are typically bounded by training data and fixed goals and are incapable of interacting with external tools or adapting to dynamic environments and new objectives. As a result, many studies that claim to use "agentic AI" in buildings are in practice limited to prompt-based question answering, where the LLM acts as a passive assistant rather than an autonomous decision-making entity [24].

- **AI Agents** represent a step forward by combining foundation models with goal-directed execution in bounded environments. They typically integrate tool calls (e.g., APIs), task-level memory, and structured actions to complete specific tasks [20][21]. Such agents can interpret context and execute short-horizon actions effectively. However, they are often designed for isolated tasks, with limited long-horizon reasoning, limited adaptability across scenarios, and weak support for multi-agent collaboration [24], which constrains their applicability in complex building ecosystems.

- **Agentic AI**, in contrast, targets end-to-end autonomy at the workflow level. Instead of following hardcoded procedures (as in traditional rule-based automation [6]), agentic AI systems can perceive context, reason about goals, generate multi-step plans, execute actions through external tools, and revise decisions using feedback-driven adaptation [7]. More importantly, modern agentic architectures often involve multiple specialized agents coordinated by an orchestrator, enabling decomposition of high-level objectives into subtasks and coordinated execution across heterogeneous modules [22][23]. Through shared memory and persistent context, these systems are particularly suitable for complex domains such as building ecosystems, where decision-making must integrate multi-scale dynamics, multiple stakeholders, and competing objectives.

In the building domain, most existing studies, as summarized in recent review papers [13][14] adopt LLMs either as prompt-based assistants for question answering or as single-agent tool-calling systems that execute predefined functions. While useful, such systems typically lack workflow-level autonomy and coordinated multi-agent decision-making, and therefore do not yet align with the definition of end-to-end agentic AI described above. As a result, the terms "agentic AI" and "multi-agent" are often used loosely in the literature, leading to conceptual ambiguity and misaligned system designs. Xu et al. [15] presented one of the first systematic multi-agent agentic frameworks in the building domain, and several other studies [16][17][18] have begun to explore related opportunities. More recently, Lee et al. [12] reviewed emerging agentic frameworks spanning building information modeling (BIM), simulation, digital twins, building management systems, and urban-scale coordination, highlighting strong potential for scalable automation. Nevertheless, substantial gaps remain before agentic AI can reliably support end-to-end intelligent building operation workflows.

- **Lack of Systematic Framework Design for Agentic AI**: Most current studies focus on single-agent applications or ad-hoc implementations for building applications. There are only a few studies on the architectural design of multi-agent coordination frameworks, specifically regarding hierarchy, structural, prompt engineering, agent development, and tool integration. Designing a robust framework that effectively orchestrates multiple agents to handle the complexity of building energy systems remains an open challenge.
- **Absence of Comprehensive Benchmark Evaluations**: There is a scarcity of studies that evaluate Agentic AI performance using systematic metrics. Existing literature lacks rigorous benchmarking regarding success rates, planning accuracy, token usage, computational cost, and execution time. This lack of standardized evaluation creates a barrier to understanding the true reliability and scalability of these systems in real-world scenarios.
- **Limited Applications due to a small toolset:** Most building-oriented agentic AI systems rely on building performance simulation engines (e.g., EnergyPlus) as the primary external tool. While effective for passive simulation and scenario analysis, these tools provide limited support for intelligent building operation workflows such as real-time control optimization and coordination across HVAC systems, DERs, and grid interactions. This constraint significantly limits the practical scope of current agentic AI solutions and motivates the need for a scalable, interactive, and physics-consistent environment that supports closed-loop decision making.

### 1.2 A Scalable, Physics-Informed Learning Environment

While agentic AI provides the "brain" for planning and decision-making, agentic AI alone is not sufficient. To produce reliable and context-aware actions, agents require a realistic environment in which they can perceive system states, execute actions, observe feedback, and adapt strategies. Direct deployment and iterative learning on physical buildings is generally infeasible due to safety risks, operational constraints, time cost, and economic burden.

Therefore, a scalable, adaptive, and physics-consistent virtual environment is essential, and it directly determines the practical capability and reliability of an agentic AI framework. However, existing modeling and simulation platforms fall short in several ways:

- **Limited Scalability and Real-Time Adaptability:** High-fidelity physics-based tools such as EnergyPlus [25] and Modelica [29] provide accurate simulation, but they are labour intensive, difficult to scale, and hard to adapt in real time to new buildings. In contrast, purely data-driven models offer scalability and speed, but often suffer from limited physical consistency and weak generalization, especially when training data are sparse, noisy, or unavailable.
- **Lack integration of Multi-Domain, Multi-Scale, Multi Component Systems:** Many existing tools focus primarily on building thermal performance and do not fully integrate DERs, grid dynamics, and occupant behavior, which are essential for future grid-interactive and multi-energy systems. While platforms such as CityLearn [30], REopt [31], and HOMER Pro [32], capture DER–grid interactions, they often treat buildings as simplified static nodes with predefined or predicted loads. This neglects the closed-loop coupling between thermal dynamics, HVAC control actions, and electric/DER operations, potentially leading to unrealistic dynamic responses and biased performance conclusions. Moreover, cross-domain coordination (e.g., thermal–electrical coupling) across HVAC and DER systems, especially under centralized or hybrid control settings is still largely unsupported.

These limitations motivate the second pillar: a Physics-Informed Machine Learning environment. PIML embeds physical knowledge (e.g., conservation laws, causal structures, and system constraints) into the learning process through model architectures, loss functions, parameterization, and training algorithms [33]. This approach enables a strong balance among accuracy, scalability, physical consistency, and generalization, and has demonstrated promising performance in various building applications. In this work, we develop a modular PIML-based runtime environment that integrates building thermal dynamics, HVAC systems, DERs, and related components, forming an interactive "body" where agentic AI can reliably learn, reason, and act.

### 1.3 Contribution and Novelty: An Agentic AI–Enabled PIML Framework

To address the above challenges, we develop a unified framework that couples a multi-agent agentic AI architecture with a modular physics-informed building energy performance modeling, control, and optimization environment (BESTOpt[26][27][28]). The main contributions of this work are summarized as follows:

- **First multi-agent agentic AI framework for intelligent building operations:** We present the first end-to-end multi-agent agentic AI framework specifically designed for intelligent building operations. The framework consists of one top-level orchestrator and 11 specialist agents, coupling LLM-based planning with a PIML environment to enable autonomous and scalable building-level decision-making.
- **Integration with BESTOpt via tool-enabled interactions:** We integrate the agentic AI framework with BESTOpt, a physics-informed modular runtime environment that incorporates building thermal dynamics, HVAC systems, DERs, grids, and occupancy. This integration is enabled through standardized Model Context Protocol tool calling with 72 domain-specific tools and supports coordinated analysis and operation at both building and cluster scales.
- **Benchmark-scale evaluation with 3975 case studies:** We conduct comprehensive case studies, generated from 53 standardized test cases spanning four workflow complexity levels (single/multi-agent × single/multi-tool), executed under 25 orchestrator–specialist model pairings and three intelligence modes. Evaluation metrics include task success rate, tool-calling accuracy, planning correctness, and computational overhead (tokens, runtime, and cost), enabling a quantitative understanding of the strengths and limitations of different agentic designs.

The remainder of the paper is organized as follows. Section 2 presents the proposed Agentic AI–PIML framework and the benchmark evaluation setup. Section 3 reports the case study findings and benchmark results. Section 4 discusses key lessons learned and implications for future agentic building systems. Finally, Section 5 concludes the paper and outlines future research directions.

## 2. Methodology

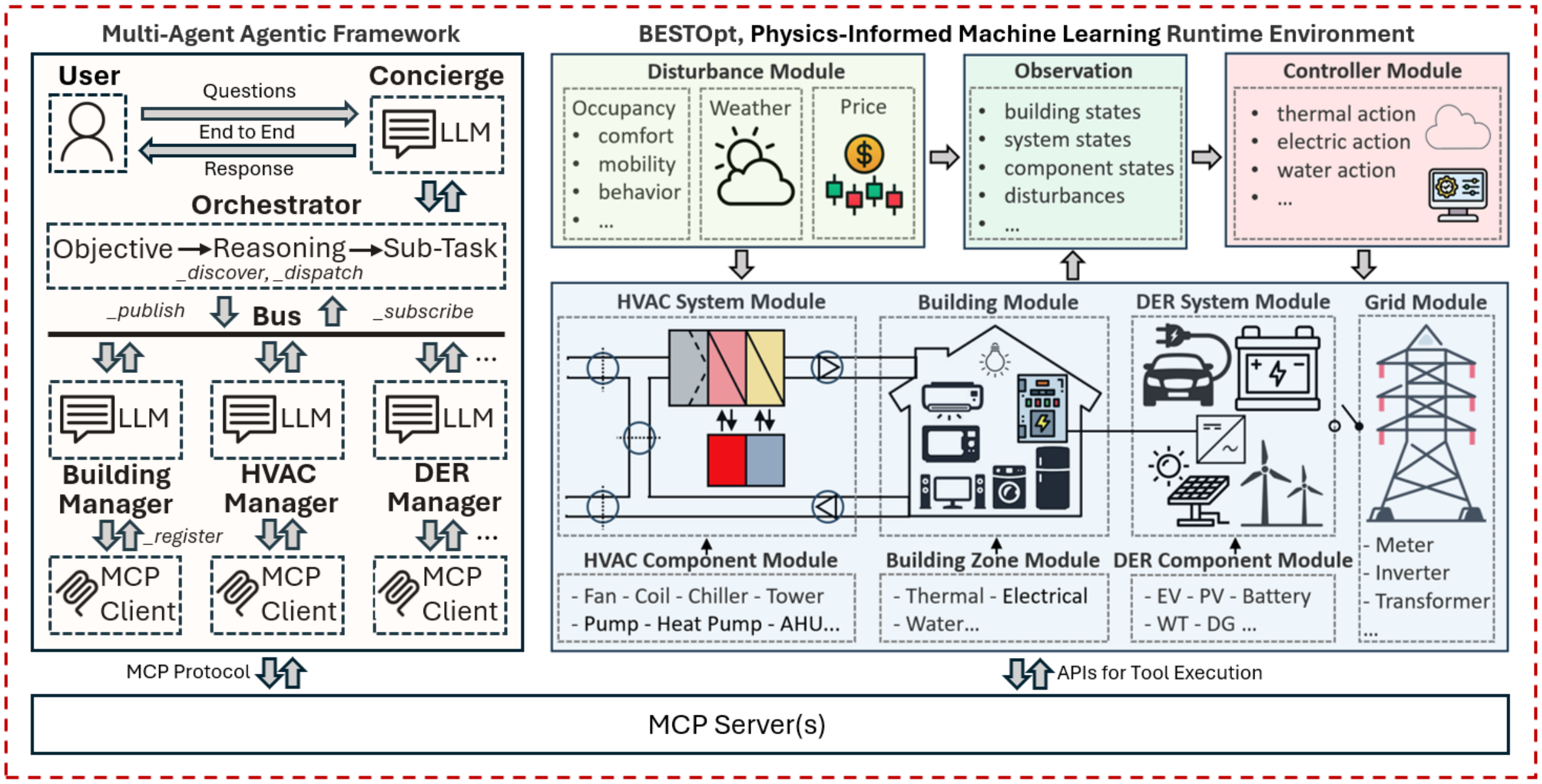

Figure 2. Overall Architecture of the Agentic AI–enabled PIML Framework for Autonomous Coordination.

Figure 2 illustrates the overall structure of the proposed Agentic AI–enabled PIML framework for autonomous occupancy–building–HVAC–DER–grid coordination. The framework is composed of two tightly coupled layers: (1) a multi-agent Agentic AI decision layer that performs goal-driven reasoning and task orchestration, and (2) a physics-informed runtime environment implemented in the BESTOpt platform that provides a realistic, adaptive, and physics-consistent simulation space for decision execution. These two layers communicate through standardized model context protocol, enabling seamless interaction between intelligent agents and the underlying physical system models.

In this section, we detail the architecture, workflow, and communication mechanisms of the proposed framework, focusing on how Agentic AI and the PIML environment jointly realize autonomous, end-to-end decision-making for complex building energy systems.

### 2.1 Multi-Agent Agentic AI Framework

The proposed Agentic AI framework is implemented as a modular multi-agent system that supports autonomous reasoning, planning, tool execution, and iterative adaptation. It adopts a hierarchical organization consisting of (1) a concierge agent for user interaction, (2) an orchestrator agent for global planning and coordination, and (3) a cluster of specialist agents for domain-specific operations (e.g., building, HVAC, DER, simulation, comparison). All agents communicate with the runtime environment through live Model Context Protocol, which exposes environment functions as standardized callable tools. This design enables coordinated workflows spanning building modeling, control configuration, simulation execution, and performance evaluation within a unified runtime loop.

### 2.1.1 Operation Workflow and Execution Logic

The system operation logic is organized into two stages: initialization and operation, as shown in Figure 3, which illustrates the end-to-end interaction pipeline between the multi-agent decision layer and the BESTOpt runtime environment. The framework first completes an initialization stage to establish the MCP tool registry and agent pool. It then enters the operation stage, where user requests are processed through a closed-loop coordination process among the concierge agent, orchestrator, and specialist agents.

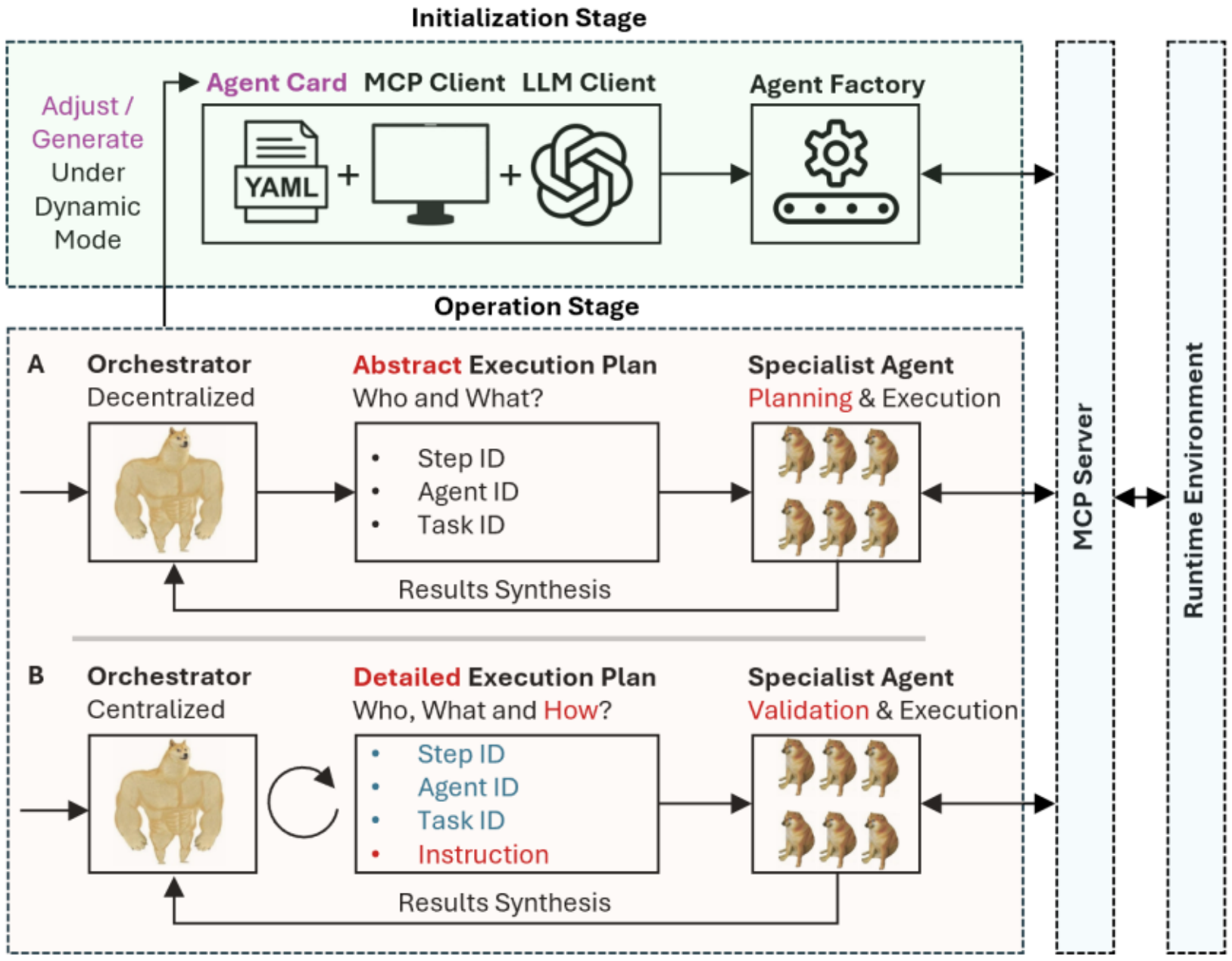

Figure 3. Initialization and Operation Workflow of the Agentic AI–enabled PIML Framework

- **Initialization Stage**: The framework first launches an MCP server to register and host the runtime environment functions as tool interfaces. An MCP client then connects to the server, queries available tools, and builds a tool index for subsequent routing. In the current implementation, we developed 72 MCP tool interfaces covering building configuration, HVAC control, DER management, simulation, analysis, comparison, and visualization. Appendix A summarizes the detailed tool information, including the tool list (Appendix I-1) and an example tool specification (Appendix I-2). Next, all LLM-based agents are instantiated via an agent factory using YAML-based agent configuration manifests (agent cards). We define 11 specialist agents in the current setup (the agent list is shown in Appendix II-1). Each agent card specifies the agent identity (name/role), capability scope, authorized tools, reasoning style, and operational constraints. This agent-card design is inspired by Google ADK-style agent configs [34], while being customized to support BESTOpt's domain decomposition and MCP tool registry. The agent card schema and a representative YAML example are shown in Appendix II-2. Once the tool registry, communication links, and agent pool are established, the framework transitions to the operation stage.

- **Operation Stage:** During operation, the framework maintains a closed-loop interaction among the user, the concierge agent, the orchestrator, and the specialist agent pool via MCP tool calls. The workflow starts with the concierge agent (details in Appendix III-1), which handles natural language communication, reformulates ambiguous instructions, and passes structured task descriptions to the orchestrator. The orchestrator functions as the central decision-maker and task manager. Specifically, the orchestrator (1) interprets the overall objective, (2) reasons over task dependencies and tool prerequisites, (3) generates a multi-step execution plan with explicit agent assignments, and (4) dispatches step-level tasks to specialist agents for execution. After all steps are completed, the orchestrator aggregates intermediate outputs, synthesizes an executive summary with key numerical findings when available, and returns the final response to the concierge agent, which presents the results back to the user.

As the core of the framework, the orchestrator workflow is illustrated in Figure 3, where we implement two intelligence modes.
**-Centralized Intelligence (Figure 3B).** A high-capability orchestrator produces a fully specified execution plan. Each plan step includes (1) the responsible agent, (2) a task description, (3) explicit step dependencies, and (4) orchestrator guidance that encodes tool-level instructions, including the selected tool(s), structured parameter fields, expected outputs, and validation checks. In our implementation, the orchestrator constructs these detailed instructions by incorporating tool schemas (parameter names, types, required flags, and semantic descriptions) into the planning prompt. However, as the number of tools and agents grows, providing full schemas for all tools and agents can substantially increase prompt length and planning complexity. To mitigate this "prompt explosion" issue, we develop **one-stage vs. two-stage centralized planning**. In one-stage planning, the orchestrator receives full descriptions and schemas for all available tools and agents and generates a detailed plan in a single pass. In two-stage planning, the orchestrator first performs a lightweight routing step using only minimal agent/tool metadata (agent identity and tool names) to produce a high-level plan that identifies the required agents and tools. In the second step, the orchestrator is provided only the schemas of the selected tools and then completes the plan by filling in tool parameters, expected outputs, and validation logic. The centralized orchestrator prompts are provided in Appendix III-2. And the related specialist agent prompt can be found in Appendix III-4.

**-Decentralized Intelligence (Figure 3A).** A lightweight orchestrator generates a high-level plan that specifies the task sequence, assigned agents, and expected outcomes, but does not prescribe detailed tool usage or parameter values. Each specialist agent then performs autonomous execution: it selects appropriate tools from its authorized tool set, configures parameters, executes MCP tool calls, and synthesizes step-level results. Compared with centralized intelligence, the decentralized plan is intentionally less prescriptive. Its primary role is to enforce global ordering and dependency structure, while tool selection and parameterization are delegated to domain agents. This mode emphasizes modularity, flexibility, and agent autonomy, particularly when tool schemas are extensive or when execution requires domain-specific heuristics. The decentralized orchestrator prompt is provided in Appendix III-3. And the related specialist agent prompt can be found in Appendix III-4.

**Dynamic agent generation and adaptation.** In addition to the two intelligence modes, the orchestrator supports an optional dynamic function (analogous to a Human Resources department that manages an organization's employees) that checks whether the current agent pool can satisfy a user request under the existing tool assignments. If capability gaps are detected (e.g., required tools are unassigned or no agent covers a requested domain), the orchestrator can (1) create a new specialist agent by generating a compliant agent card with an appropriate tool subset, or (2) revise an existing agent card by expanding its authorized tool list. This mechanism improves robustness under evolving user needs and expanding tool registries, enabling the framework to adapt without manual redesign of the initial agent pool. The detailed prompt can be found in Appendix III-5.

### 2.1.2 Communication and Coordination Mechanisms

The proposed framework adopts a layered communication architecture to enable coordinated interactions among the user, LLM agents, and the BESTOpt runtime environment. This architecture includes three complementary components: (1) an Agent-to-Agent (A2A) communication protocol for collaboration among specialist agents, (2) the Model Context Protocol for structured agent–environment interaction, and (3) a shared BUS that serves as the runtime messaging backbone. Conceptually, A2A and MCP specify *what* entities communicate and *how* messages are structured, while the BUS provides the underlying channel through which messages are delivered, routed, and logged.

#### 2.1.2.1 A2A Communication

A2A communication supports coordination among specialist agents responsible for distinct subtasks (e.g., system configuration, simulation, and performance evaluation). Through A2A messaging, agents exchange intermediate results, task status, and structured summaries, enabling collaborative problem solving without requiring a single agent to maintain full domain coverage. In our design, A2A interactions are explicitly scoped and task-driven, which helps maintain transparency, traceability, and predictable system behavior.

#### 2.1.2.2 Agent–Environment Interaction via MCP

Interactions between agents and the BESTOpt runtime environment are standardized through MCP, which provides a unified tool-based interface for accessing environment functionalities. Specifically, environment functions are packaged as MCP-compliant tools using the FastMCP toolkit [35]. Each tool is defined by a standardized name and description, together with a machine-readable input schema and output format. Tools are hosted on an MCP server, while MCP clients embedded within agents support dynamic tool discovery and invocation. Through MCP tool calls, agents can execute control actions, query system states, trigger simulations and optimization routines, and retrieve performance feedback. This design enforces a clean separation between (1) agent intelligence (reasoning, planning, and decision-making) and (2) physical system execution (simulation and model-based computation), improving modularity and reproducibility.

#### 2.1.2.3 BUS as the Runtime Communication Medium

The BUS functions as the shared runtime messaging layer that enables communication across system components, including the user interface, concierge/orchestrator agents, specialist agents, and the BESTOpt environment. It supports message passing for task requests, tool invocation events, execution updates, intermediate outputs, and system-level status notifications. By providing a unified communication channel, the BUS reduces tight coupling between individual agents and services, thereby improving scalability and maintainability. Within the agentic layer, the BUS enables A2A coordination, while also carrying MCP-based requests and responses exchanged between agents and the environment.

## 2.2 BESTOpt: A Modular, Physics-Informed Machine Learning based Building Modeling, Control and Optimization Framework

As shown in the right panel of Figure 2, BESTOpt serves as the physics-grounded runtime environment ("body") of the proposed Agentic AI–PIML framework. It provides a modular and physics-informed platform for integrated modeling, control, and optimization of multi-domain building energy systems, including occupancy behavior, building thermal dynamics, HVAC operation, DERs, and grid interactions. By enabling closed-loop execution (state perception → action application → environment feedback), BESTOpt offers a scalable testbed for evaluating and deploying autonomous workflows in resilient, energy-efficient, and grid-responsive building operation. Interested readers are referred to our prior work[26][27][28] for full BESTOpt details.

In this paper, to support understanding of the agent–environment interaction, BESTOpt is briefly summarized through three key design mechanisms as shown in Figure 4:

- **Hierarchical modular structure:** BESTOpt adopts a scalable **cluster–domain–system/building–component** hierarchy. A cluster defines the overall coordination boundary (e.g., neighborhood, campus, district) and may include an arbitrary number of buildings, HVAC systems, and DER assets. An intermediate domain layer organizes interactions by energy pathways (e.g., thermal, electrical, and water domains). Within each domain, BESTOpt represents physical entities as systems/buildings (e.g., HVAC plants, DER systems, individual buildings), which are further decomposed into controllable components (e.g., coils, fans, batteries, EVs), enabling both system-level coordination and device-level control within a unified structure.

- **Unified data typology and flow:** BESTOpt standardizes runtime information exchange using a unified **state–action–disturbance–observation** representation. States describe internal system dynamics (e.g., zone temperature, battery SOC), actions represent control decisions (e.g., setpoints, dispatch commands), disturbances capture exogenous drivers (e.g., weather, prices, occupancy), and observations include measured/inferred/predicted information used for monitoring and decision-making. This design enables a structured bidirectional data flow, where component-level variables are aggregated upward for system/cluster reasoning, while high-level objectives and decisions are decomposed downward into device-level commands.
- **Module types and execution logic:** BESTOpt integrates three categories of modules: dynamic modules, controller modules, and disturbance modules. All modules follow a unified execution interface (*initialize–reset–step*), and are synchronized at each simulation timestep. Specifically, disturbance modules update exogenous signals, controller modules compute actions from observations and upstream objectives, and dynamic modules update system states accordingly, which enable BESTOpt to function as a live, interactive runtime environment for closed-loop autonomy.

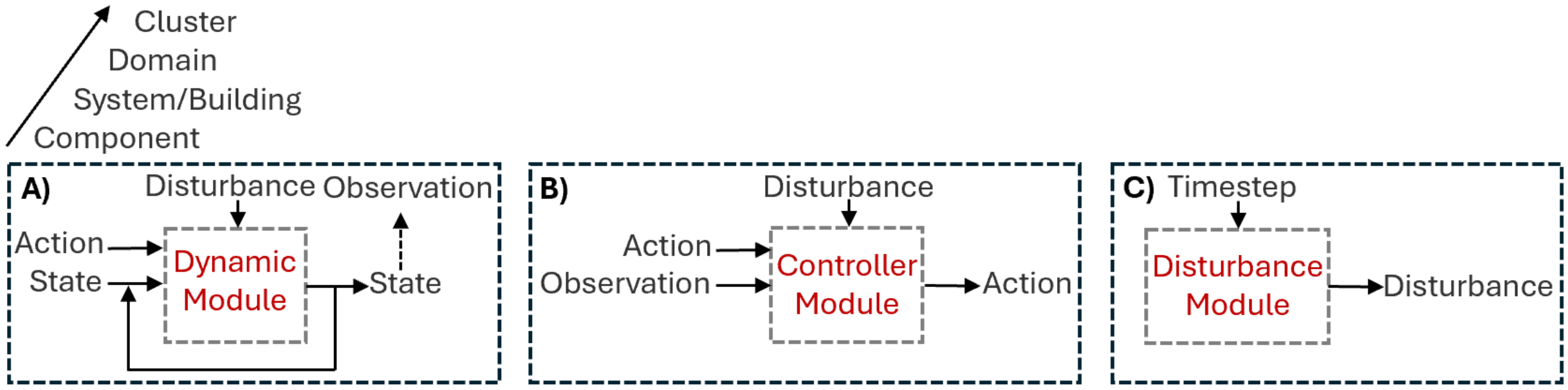

Figure 4 Modular hierarchy and data typology of BESTOpt runtime environment

### 2.3 Case Study and Evaluation

To demonstrate the capabilities and validate the performance of the proposed multi-agent automation framework, this section presents (1) a representative end-to-end case study and (2) a large scale systematic benchmark evaluation. The case study provides a detailed walkthrough of how the framework coordinates multiple specialist agents to answer a complex user query. The benchmark evaluation further quantifies framework performance under diverse configurations, including different model capability tiers, intelligence modes, and task complexities..

#### 2.3.1 Example of an End-to-End Case Study

We first present a representative end-to-end case study to illustrate the execution trace and demonstrate the practical capability of the proposed framework. In this scenario, a building energy manager asks:

“How do energy use, operating cost, thermal comfort, and flexibility metrics change if the HVAC COP is upgraded to 4.5 and the battery capacity is increased to 20 kWh for the reference test building? Furthermore, under this upgraded configuration, what is the additional impact of applying a pre-cooling strategy that lowers the zone setpoint by 2°C for 2 hours?”

This query is representative because it involves cross-domain system upgrades and requires coordinated, multi-step reasoning and tool execution. First, it needs **multi-domain coordination** across both the thermal domain (HVAC

Coefficient of Performance (COP) upgrade) and the electrical domain (battery replacement), which requires collaboration between the hvac_agent and der_agent under the orchestrator's control. Second, the question implicitly requires a **comparative evaluation workflow** rather than a single simulation run: the framework must preserve the baseline configuration, execute a baseline simulation, apply parameterized upgrades, conduct an upgraded simulation, and then compute baseline-versus-upgrade deltas. Third, the requested outputs span **multi-dimensional performance reporting**, including energy, cost, comfort, and flexibility, requiring the comparison_agent to consolidate results from multiple analysis tools into a unified report. Finally, the query imposes **parameter extraction and mapping** requirements: numerical values must be correctly parsed from natural language, validated, and mapped to the corresponding configuration fields and control logic in the runtime environment. The quantitative results from this case study are presented in the Results section.

#### 2.3.2 Benchmark Evaluation Design

Beyond the representative case study, a rigorous evaluation requires a benchmark suite that systematically tests framework performance across diverse scenarios. We develop a benchmark consisting of 50 distinct test cases, executed across 25 model configurations and three intelligence modes, resulting in approximately 4,000 individual runs. In the following subsections, we aim to asnwer three key methodological questions: How are the test cases be categorized? How are standardized test cases constructed with ground-truth expectations for evaluation? And what metrics are used to measure accuracy, efficiency, and resource usage?

##### 2.3.2.1 Benchmark Configuration

The benchmark configurations are defined along three dimensions: **test case taxonomy**, **LLM model configuration**, and **intelligence mode**.

**(1) Test Case Taxonomy**

The benchmark test cases are organized along two orthogonal dimensions: agent complexity (single vs. multi-agent) and tool complexity (single vs. multi-tool). This 2×2 taxonomy yields four categories of increasing difficulty:

- **Single-Agent Single-Tool (SAST)**: A single specialized agent invokes one tool to complete the request. These cases establish baseline performance for fundamental operations such as creating configurations, querying system parameters, or adding individual components.
- **Single-Agent Multi-Tool (SAMT)**: A single agent orchestrates multiple tool invocations in sequence. These cases evaluate an agent's ability to maintain context across sequential operations within a single domain, such as adding a system, updating its parameters, and querying the final result.
- **Multi-Agent Single-Tool (MAST)**: Multiple specialized agents coordinate, each invoking a single tool. These cases assess the orchestrator's inter-agent coordination capabilities, such as validating a configuration before running a simulation, or adding an HVAC system followed by its controller.
- **Multi-Agent Multi-Tool (MAMT)**: Multiple agents coordinate with each potentially invoking multiple tools. These cases represent realistic complex requests and evaluate end-to-end framework performance.

**(2) LLM Model Configuration**

To study the relationship between model capability and framework performance, we define a model capability matrix spanning five tiers, ranging from commercial API-based models to lightweight locally deployed models. As shown in Table 2, the benchmark evaluates all 25 pairwise combinations of orchestrator and specialist-agent model assignments.

Table 2 LLM Model Tiers Configuration

| Tier | Model | Parameters | Deployment | Price |
|---|---|---|---|---|
| API-class | gpt-4o-mini | - | Cloud API | Input/Output: $0.15 / $0.60 per 1M tokens |
| Extra-Large (XL) | qwen3:30b | 30B | Local (Ollama) | Free |
| Large (L) | gemma2:9b | 9B | Local (Ollama) | Free |
| Medium (M) | qwen3:4b | 4B | Local (Ollama) | Free |
| Small (S) | qwen3:1.7b | 1.7B | Local (Ollama) | Free |

**(3) Intelligence Mode Configuration**

Each benchmark case is executed under three intelligence modes to compare centralized versus decentralized coordination strategies as shown in Table 3.

Table 3 Intelligence Mode Configuration

| Mode | Label | Description |
|---|---|---|
| Centralized Single-Stage | C-1 | Orchestrator generates complete execution plan with tool-level instructions in one pass |
| Centralized Two-Stage | C-2 | Orchestrator first generates high-level plan, then elaborates with detailed tool instructions |
| Decentralized | D | Orchestrator provides goal-level planning; agents autonomously select tools and parameters |

#### 2.3.2.2 Benchmark Test Case Structure

Each benchmark test case is defined as a structured object containing the user request, expected execution behavior, and ground truth for accuracy evaluation as shown in Table 4.

Table 4 Test Case Schema Definition

| Field | Type | Description |
|---|---|---|
| test_id | string | Unique identifier encoding category and domain (e.g., "MAMT_001") |
| name | string | Human-readable test case name |
| category | enum | Complexity category: SAST, SAMT, MAST, or MAMT |
| request | string | Natural language user query |
| expected_agents | list[string] | Set of agents that should be invoked |
| expected_tools | list[string] | Set of tools that should be called |
| expected_steps | list[ExpectedStep] | Ordered sequence of execution steps |
| description | string | Brief description of test purpose |

Each ExpectedStep further specifies:

- **step_order**: Sequential position in execution plan
- **agent_id**: The specialist agent responsible for this step
- **required_tools**: Tool(s) the agent should invoke
- **expected_parameters**: Ground truth parameter values for each tool call

#### 2.3.2.3 Benchmark Evaluation Metrics

The benchmark records metrics across four dimensions: accuracy, timing, resource usage, and cost as shown in Table 5.

Table 5 Test Case Performance Indicator Metrics

| **Dimension** | **Metric** | **Definition** |
|---|---|---|
| Accuracy | Tool selection accuracy | Jaccard similarity between expected vs. executed tool sets |
| Accuracy | Agent selection accuracy | Ratio of correctly engaged agents vs. expected agents |
| Accuracy | Plan step accuracy | Alignment between expected and executed step order |
| Accuracy | Parameter accuracy | Correctness of extracted argument keys and values (including numeric match) |
| Timing | Total time | End-to-end request processing duration |
| Timing | Planning time | Orchestrator planning latency |
| Timing | Execution time | Aggregated agent tool execution duration |
| Timing | Synthesis time | Response synthesis duration |
| Tokens | Orchestrator tokens | Prompt + completion tokens consumed by orchestrator |
| Tokens | Agent tokens | Total prompt + completion tokens across all agents |
| Tokens | Total tokens | Combined token usage (orchestrator + agents) |
| Cost | Orchestrator cost | Inference cost for orchestrator operations |
| Cost | Agent cost | Aggregate inference cost for agent operations |
| Cost | Total cost | Total cost per request |

**Accuracy metrics.** For each test case, the benchmark defines ground-truth expectations including the required tools, expected agents, step sequence, and tool arguments by $T_{\text{exp}}$, $G_{\text{exp}}$ and $S_{\text{exp}}$. We extract the executed tool set $T_{\text{act}}$, executed agent set $G_{\text{act}}$, and executed step trace $S_{\text{act}}$from persisted per-agent execution logs. $Acc_*$ represents the accuracy of tool, agent, step and parameters.

Tool selection accuracy is computed as expected-set recall:

$$Acc_{tool} = \frac{|T_{exp} \cap T_{act}|}{|T_{exp}|} \quad \textit{Equation 1}$$

Where $|\cdot|$ is cardinality set size and $T_{exp} \cap T_{act}|$ means the number of tools correctly executed. Similarly, Agent selection accuracy is computed by:

$$Acc_{agent} = \frac{|G_{exp} \cap G_{act}|}{|G_{exp}|} \quad \textit{Equation 2}$$

Plan step accuracy measures order-consistent execution. Let the expected step sequence be $S_{\text{exp}} = \{(g_i, \mathcal{T}_i)\}_{i=1}^{N}$and executed steps be $S_{\text{act}} = \{(\hat{g}_j, \hat{\mathcal{T}}_j)\}_{j=1}^{M}$. $g_i$ and $\mathcal{T}_i$ represents tool use and agent selection per step. Using subsequence matching, an expected step is considered matched when $\hat{g}_j = g_i$and $\mathcal{T}_i \subseteq \hat{\mathcal{T}}_j$in chronological order. If $K$out of $N$ steps are matched, then:

$$Acc_{plan} = \frac{K}{N} \quad \textit{Equation 3}$$

Parameter accuracy is evaluated at both the argument-key and argument-value levels by comparing expected vs. extracted tool arguments. Key accuracy is the fraction of expected keys that appear in the best-matching executed call; value accuracy is the fraction whose values match under type-aware comparison.

$$Acc_{key} = \frac{|K_{exp} \cap K_{act}|}{|K_{exp}|} \quad \textit{Equation 4}$$

$$Acc_{val} = \frac{|V_{exp} \cap V_{act}|}{|V_{exp}|} \quad \textit{Equation 5}$$

**Timing metrics.** Latency is recorded using orchestrator timestamps. Total time measures end-to-end duration from request submission to final response. Planning, execution, and synthesis time correspond to orchestrator planning, specialist tool execution, and response generation stages, respectively.

**Token and cost metrics.** Tokens are obtained from provider usage records for each LLM call and aggregated by role: orchestrator tokens (prompt and completion across orchestrator calls) and agent tokens (sum across specialists).. Cost is computed by multiplying token usage by the corresponding per-token pricing for each model/provider and summing across calls, including orchestrator cost, agent cost, and total cost.

In summary, The benchmark suite implements checkpointing with SQLite-based persistence, enabling interruption and resumption of long-running evaluation campaigns. Each test configuration is uniquely identified by a composite key comprising test case ID, intelligence mode, model configuration, and repetition index. Completed results are cached to avoid redundant computation during resumed runs.

## 3. Results

### 3.1 Case Study of System Upgrade Impact Assessment

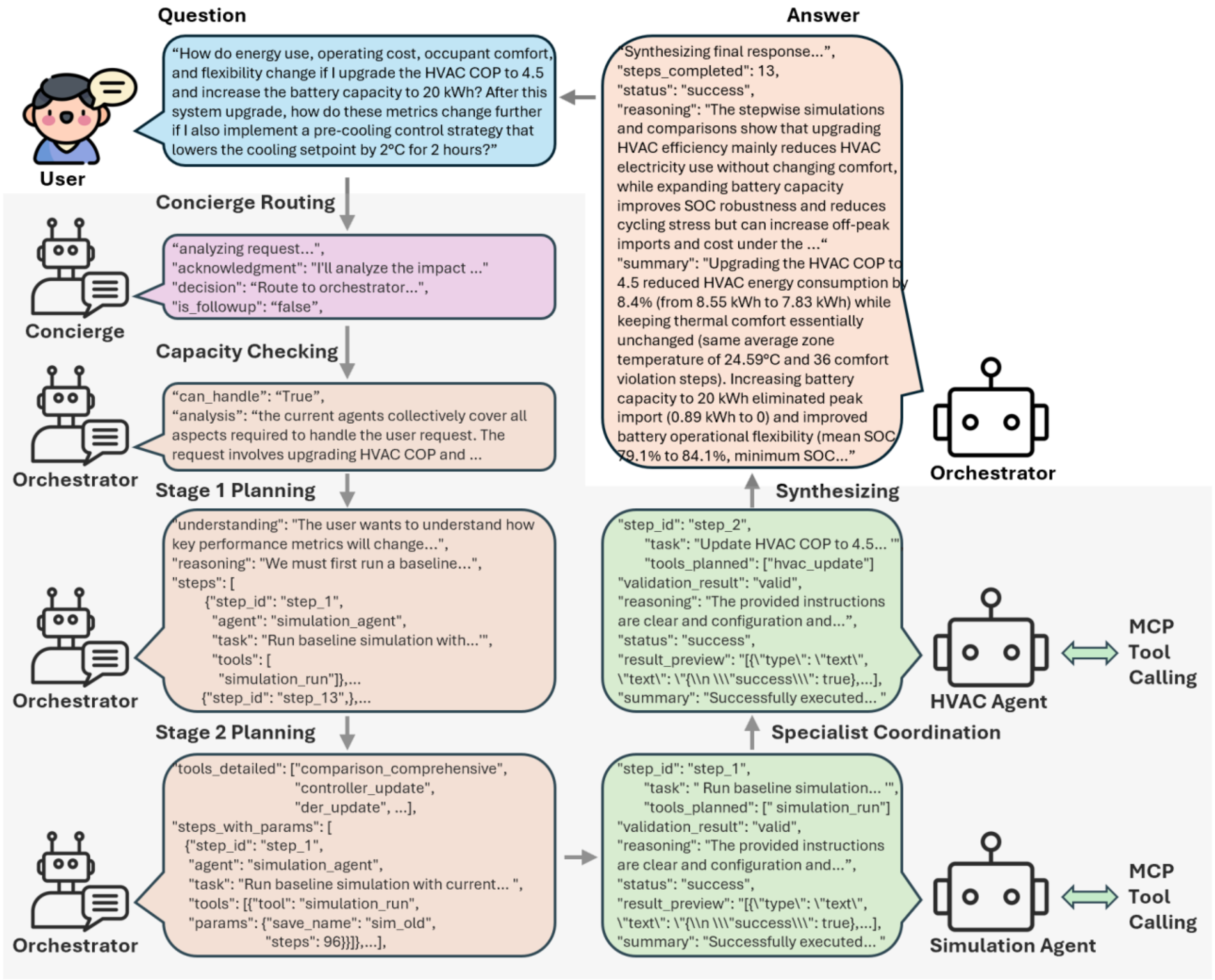

Figure 5. Reasoning Trace for an Example Case Study

This section demonstrates an end-to-end case study to illustrate how the proposed agentic AI framework executes a realistic building energy management query as shown in Figure 5. To address this multi-step request, the framework autonomously conducts three sequential simulations: (1) Baseline configuration, (2) System upgrade, and (3) System upgrade with control upgrade. Figure 5 presents a representative reasoning trace, illustrating the orchestration process under the configuration of GPT-4o-mini, centralized intelligence, and two-stage planning with dynamic model routing enabled. Once the concierge agent receives the user query, it preprocesses the request and forwards it to the orchestrator. The orchestrator first verifies tool and agent availability under the current runtime configuration, then generates a high-level execution plan specifying the required specialists, tool calls, and simulation steps. Next, a second-stage plan refinement step expands this plan using detailed agent card metadata and tool schemas for only the selected agents/tools, yielding an executable workflow with explicit parameters. The finalized execution plan is dispatched to the specialist agents, which interact with the PIML environment via MCP tool interfaces and return intermediate results after each step. The framework then aggregates and summarizes the simulation outputs into the final response.

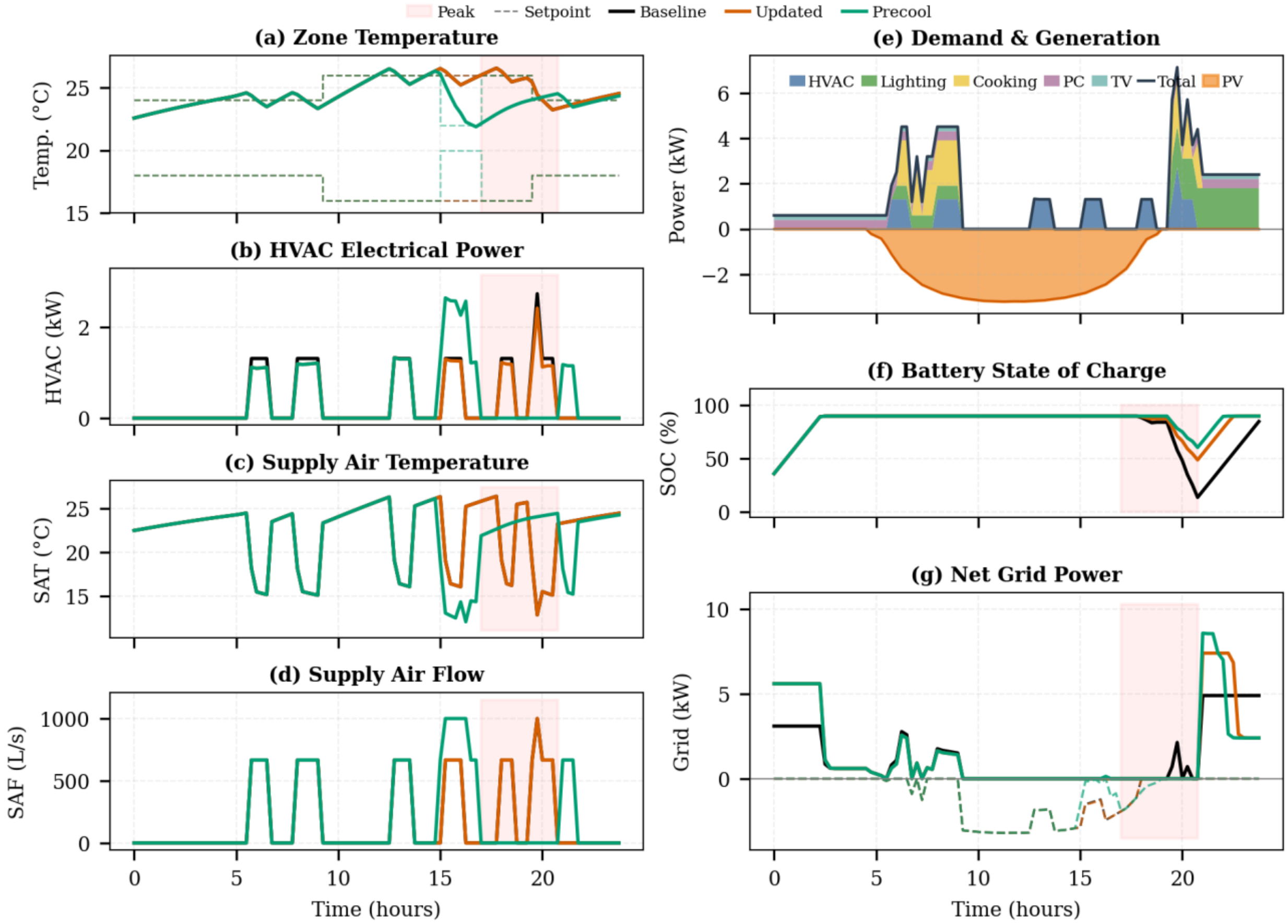

Figure 6. Performance Assessment for a Representative Cooling Day (August 1st) with system (HVAC COP and battery capacity) and controller (precooling) upgrade. The baseline case is shown in black, the system-upgrade case in red, and the system with control upgrade case in green. Peak hours are highlighted using a shaded region.

- **Thermal-domain results (Figure 6a–d):** (1) Thermal comfort: The baseline and COP-upgraded cases show similar zone temperature trajectories because COP improvement modifies HVAC efficiency but does not directly change the setpoints or control logic. In contrast, precooling reduces the total number of temperature violation steps, and the temperature standard deviation decreases by 4.4%. (2) Cooling energy: The COP upgrade reduces HVAC energy consumption by 8.4%. With precooling enabled, total cooling demand slightly increases due to intentional load shifting ahead of peak hours, reflecting a flexibility-driven tradeoff.
- **Electrical-domain performance (Figure 6e–g):** (1) PV utilization: Overall PV generation remains the same, but utilization patterns change significantly. Under the baseline controller, PV curtailment is high (22.56 kWh), suggesting a limited ability to absorb midday solar. Battery upgrade alone does not reduce curtailment, indicating that increasing capacity without redesigning control policies does not guarantee improved renewable utilization. (2) Flexibility via precooling: Precooling reduces curtailment to 20.26 kWh (−10.2%) and improves PV self-consumption by 20.8%, since increased midday cooling demand provides an additional sink for PV energy. (3) Battery operation: Battery upgrade improves buffering capability, with the minimum SOC increasing by 22.2%, and reduces cycling stress (equivalent full cycles: 1.01 → 0.68). With precooling, battery cycling decreases further (−44.2% vs. baseline), as more energy balancing occurs through direct PV-to-load consumption and load shifting. (4) Grid import and cost: Both upgraded cases eliminate peak grid import, but total grid import increases (battery upgrade: +21.3%) due to schedule-based charging behavior. Precooling partially mitigates this effect, but grid import remains 16.0% higher than baseline. Correspondingly, total daily cost rises by 19.5% under system upgrade and by 14.3% with precooling, since the rule-based DER control limits economic gains despite flexibility improvements, motivating the need for more advanced coordinated control policies in future studies.

## 3.2 Results of Benchmark Evaluation

### 3.2.1 Impact of Intelligence Mode on Benchmark Performance

Figure 7 quantifies how intelligence mode design affects benchmark performance in terms of accuracy, token usage, runtime, and inference cost. Overall, the centralized two-stage mode provides the most balanced performance, achieving the highest reliability while maintaining low execution latency and inference cost.

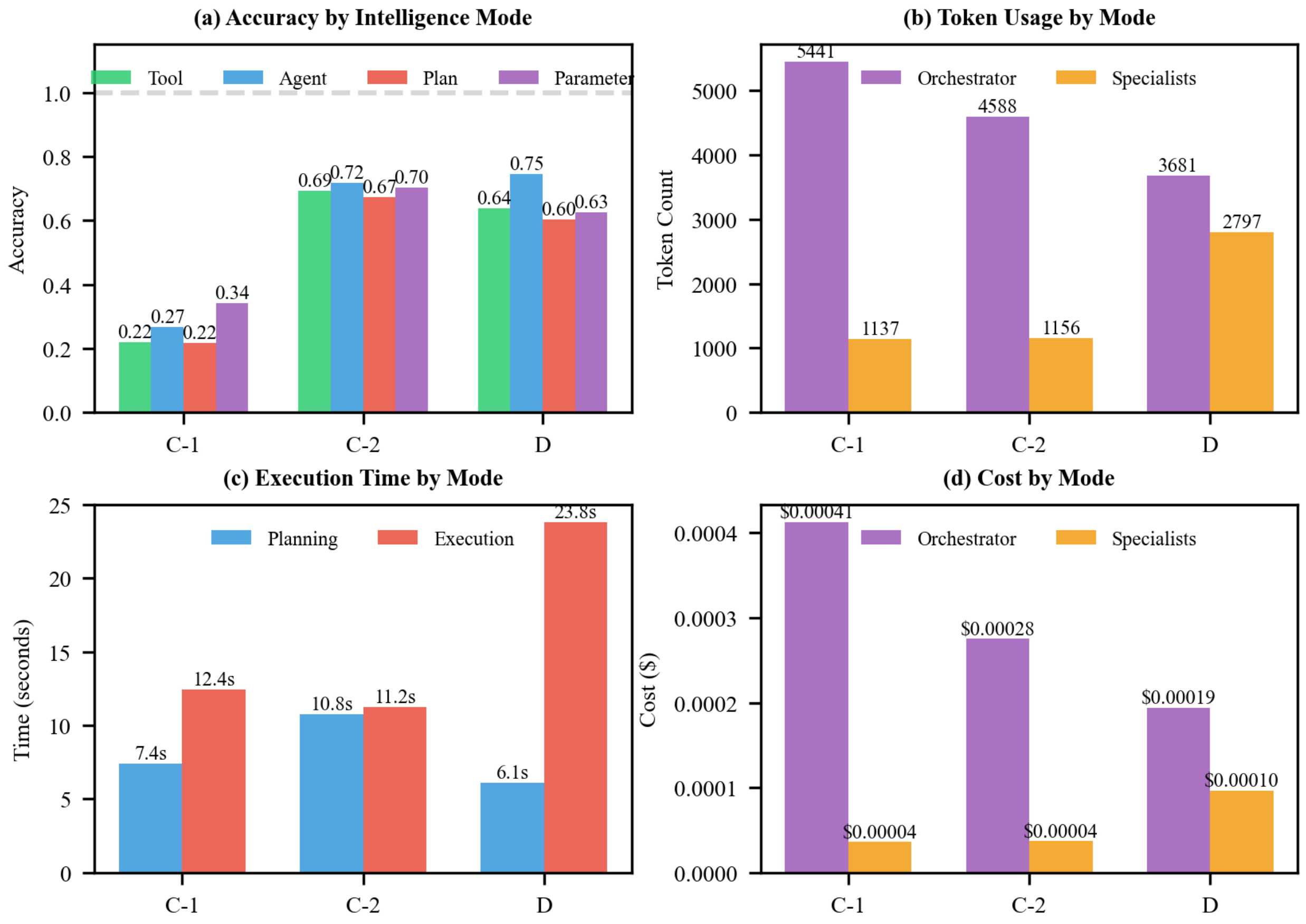

Figure 7. Impact of Intelligence Mode on Benchmark Performance. (a) Accuracy comparison across tool selection, agent selection, plan quality, and parameter extraction under centralized single-stage (C-1), centralized two-stage (C-2), and decentralized (D) modes. (b) Token consumption split between orchestrator and specialist agents. (c) Runtime decomposition into planning and execution. (d) Inference cost breakdown for orchestrator and specialists. Values are averaged over the benchmark test suite.

- **Accuracy:** As shown in Figure 7 (a), C-2 consistently outperforms C-1 across all accuracy metrics, improving tool selection accuracy to 0.69, agent selection accuracy to 0.72, planning accuracy to 0.67, and parameter identification accuracy to 0.70, compared with C-1 (0.22/0.27/0.22/0.34). These results suggest that structured two-stage reasoning is effective in mitigating prompt explosion and is particularly beneficial for preventing early-stage routing and sequencing errors, which are often the dominant failure sources in multi-step agentic workflows. Compared with decentralized mode D (0.64/0.75/0.60/0.63), C-2 still achieves slightly higher accuracy in three out of four workflow-critical categories (tool selection, planning, and parameter identification).

- **Token usage:** Figure 7 (b) shows that C-2 reduces orchestrator token usage by approximately 16% (from 5441 to 4588) relative to C-1, while specialist token usage remains nearly unchanged (1137 vs. 1156). In contrast, decentralized mode D further reduces orchestrator tokens by approximately 32% compared with C-1 (to 3681), but introduces substantial specialist overhead: specialist token usage increases by approximately 146%. Therefore, decentralization primarily shifts the reasoning burden from the orchestrator to specialists, leading to comparable total token usage while increasing agent-side deliberation overhead.
- **Runtime:** Figure 7 (c) indicates that C-2 increases planning time (~3.4s) compared with C-1 due to the additional reasoning stage in the two-stage planning schema. Decentralized mode D exhibits the shortest planning time because the orchestrator only generates high-level tasks. However, D incurs the largest execution-time overhead, 112.5% higher than C-2, because specialist agents must independently conduct additional rounds of decision-making and validation during tool execution. Overall, D results in approximately 36% higher total runtime than C-2, suggesting that decentralization shifts complexity from upfront planning into downstream execution, where it becomes more expensive and time-consuming.
- **Cost:** Figure 7 (d) C-2 reduces the total inference cost by approximately 28% compared with C-1, primarily due to reduced orchestrator token usage. While decentralized mode D achieves the lowest orchestrator cost (53.6% lower than C-1 and 32.1% lower than C-2), it more than doubles specialist-side cost, offsetting much of the benefit. This again confirms that decentralization reduces centralized coordination cost at the expense of increased reasoning overhead at the specialist level.

These results support C-2 as the default operating mode for scalable evaluation, as it achieves a stronger tradeoff between correctness, efficiency, and cost compared with fully decentralized autonomy.

### 3.2.2 Impact of Model Configuration on Benchmark Performance

Figure 8 presents the capability matrix of the proposed multi-agent framework across 25 model-size combinations (5 orchestrator tiers × 5 specialist tiers, from S→M→L→XL→API) under three intelligence modes (D, C-2, and C-1, shown from top to bottom within each cell). Each matrix cell corresponds to one orchestrator–specialist pairing and reports the average combined accuracy over the benchmark suite, aggregated from planning, agent selection, tool selection, and parameter extraction. This visualization enables a systematic examination of how performance scales with orchestrator strength, the marginal gains from scaling specialist agents, and the effectiveness of asymmetric configurations (e.g., strong orchestrator with lightweight specialists versus weak orchestrator with strong specialists).

- **Orchestrator Capability Dominates System Performance:** Across all 25 pairings, benchmark accuracy is primarily driven by the orchestrator tier, rather than specialist tier. Upgrading the orchestrator yields substantial gains even when specialist agents remain small. For instance, with the strongest orchestrator (API), combined accuracy remains consistently high, even for the worst one stage centralized configuration as discussion above section, which can still reach over 90% accuracy across matrixes. In contrast, scaling specialist agents while keeping the orchestrator weak yields limited improvement. This is because the orchestrator governs workflow-critical decisions, including task decomposition, agent/tool routing, and step sequencing. Once these upstream decisions fail, specialist capability becomes largely irrelevant, as specialists may be assigned incorrect subtasks, invoke inappropriate tools, or execute steps in an invalid order. This observation provides actionable guidance for system deployment. (1) When computational resources are sufficient, the framework becomes less sensitive to model combination, (2) when equipped with a strong orchestrator, lightweight specialists can achieve similar near-optimal performance compared to other sizes of models, (3) conversely, under constrained budgets, allocating the most capable model to the orchestrator role yields the largest and most reliable performance return.

- **Larger LLM models do not necessarily yield higher reliability:** An additional finding is that scaling model size does not monotonically improve performance. Specifically, the XL tier performs slightly worse than L in several configurations, and the M tier exhibits the weakest overall reliability. For example, the 4B model shows a higher rate of complete failures (49.3% vs. 45.2% zero-accuracy cases) and produces fewer perfect executions (46.2% vs. 50.9%) compared with a small size model (1.7B). This gap becomes more pronounced in multi-agent coordination tasks: M achieves only 44.2% tool selection accuracy compared to 57.5% for S model. The potential explanation is that smaller models tend to behave as naive executors: they strictly follow instructions, which aligns well with deterministic benchmark workflows. Larger models behave as expert planners: they can reason over multi-step structure and remain logic and structure. Mid-sized models fall in between that attempt planner-like behavior without sufficient intelligence, in other words, "mean well but do harm", thereby producing more workflow drift (e.g., incorrect tool calls, unnecessary steps, or incorrect sequencing). Regarding the XL–L comparison, two additional factors may contribute. First, extra-large models can be more fragile under constrained compute resources. Under heavy load, XL models (30B) are more prone to timeouts, crashes, or incomplete responses, introducing failure modes unrelated to reasoning quality, whereas L-tier models (9B) are typically more stable. Second, the observed difference may also reflect model-family effects (e.g., Gemini vs. Qwen) rather than size alone. A more comprehensive benchmark across model families and task types, particularly those requiring implicit reasoning beyond structured tool execution would be necessary to isolate these effects.

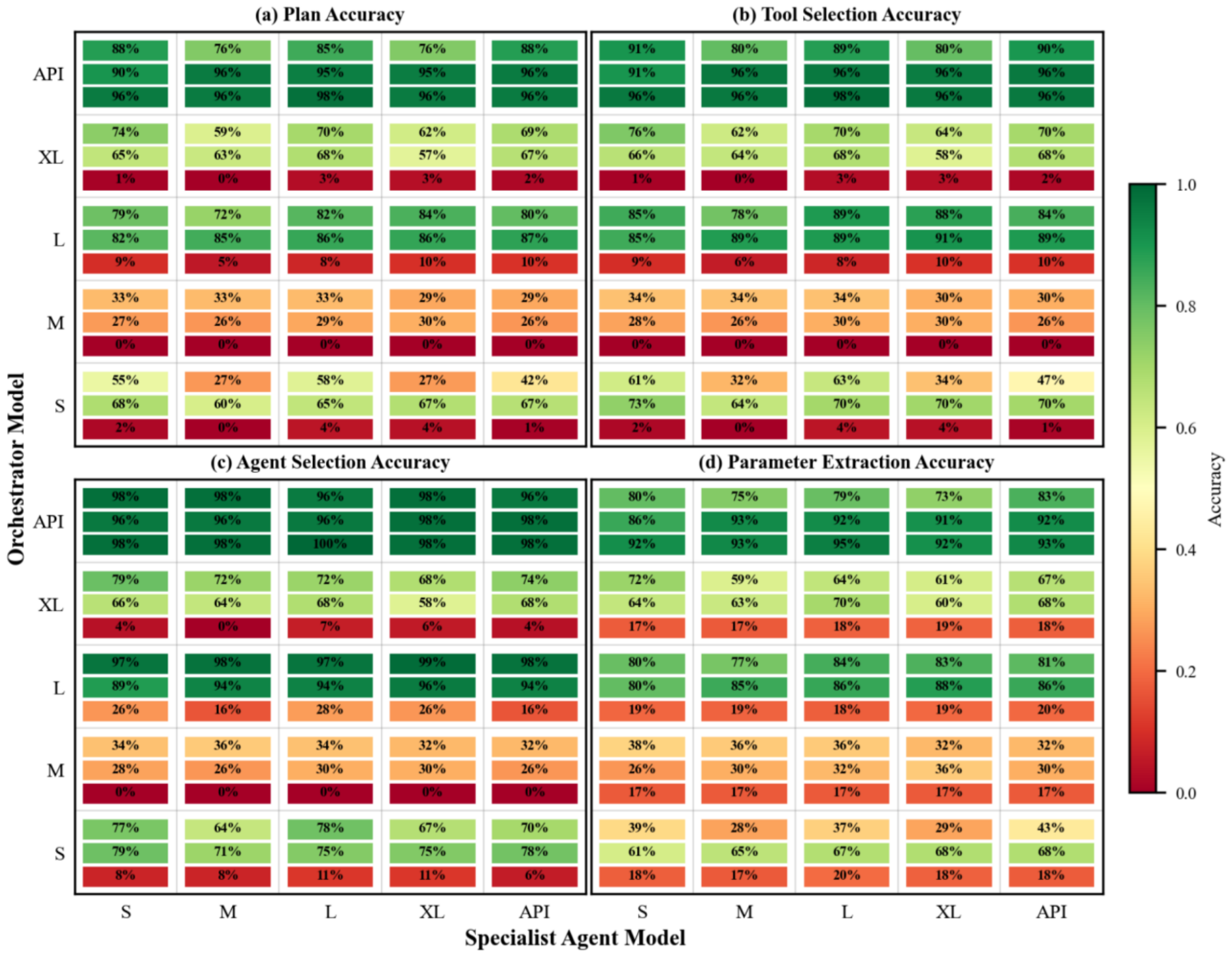

Figure 8. Impact of Model-Size Pairing on Multi-Agent Benchmark Capability.

### 3.2.3 Impact of Task Complexity on Benchmark Performance

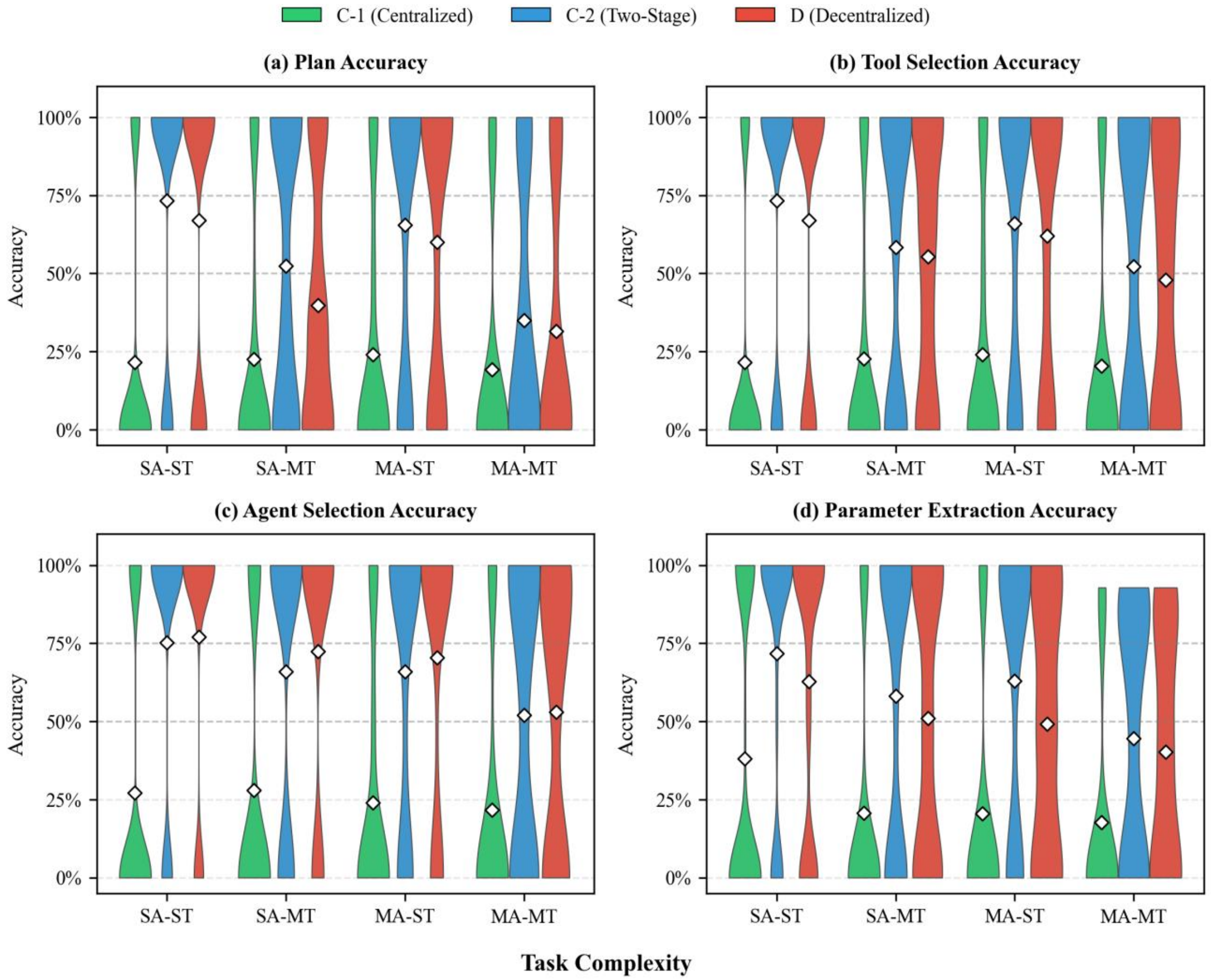

Figure 9. Impact of Task Complexity on Multi-Agent Benchmark Capability.

- **"All-or-nothing" execution behavior in agentic workflows:** Figure 9 summarizes the effect of task complexity on benchmark performance using violin plots across four accuracy metrics. Overall, the results demonstrate a clear bimodal distribution: across all accuracy metrics, most runs either achieve near-complete success or substantial failure. Specifically, 50.0% of all experiments obtain over 70% accuracy, while 46.1% fall below 30%, leaving only 3.8% in the intermediate range. Among different configurations, API-tier remain robust across all complexity levels, sustaining 91–95% accuracy, while several local model tiers degrade significantly as task complexity increases. For instance, the XL tier collapses from 53.9% to 0.0% on complex tasks. Notably, the smallest S-tier shows the least degradation (18.7%) and can outperform larger local models, consistent with the reliability trends discussed in Section 3.2.2. This bimodal distribution pattern also suggests a strong "all-or-nothing" execution behavior in agentic workflows: when the orchestrator correctly interprets the request, assigns the appropriate specialists, and sequences tool calls properly, the workflow tends to succeed end-to-end; conversely, early-stage planning errors often propagate across subsequent steps and lead to near-total breakdown. Highlighting that we must setup a strict standard threshold for agentic workflow for real world engineering problems.
- **Multi-tool usage is the dominant bottleneck compared to multi-agent coordination**: For SA–ST tasks, 53.9% of model configurations achieve high execution accuracy, but this proportion decreases to 25.8% for MA–MT tasks. Importantly, the sharp decline is primarily driven by multi-tool reasoning difficulty. While specialist agents can often be selected reliably using role definitions and agent cards, correct tool usage requires mapping ambiguous natural-language requests into structured tool arguments, selecting the correct function among similar tools, and maintaining consistent parameter extraction (key and values) across steps. These requirements significantly increase failure rates and explain the observed performance collapse under MA–MT settings.

#### 3.2.4 Impact of Orchestrator-Specialist Agent Coordination on Benchmark Performance

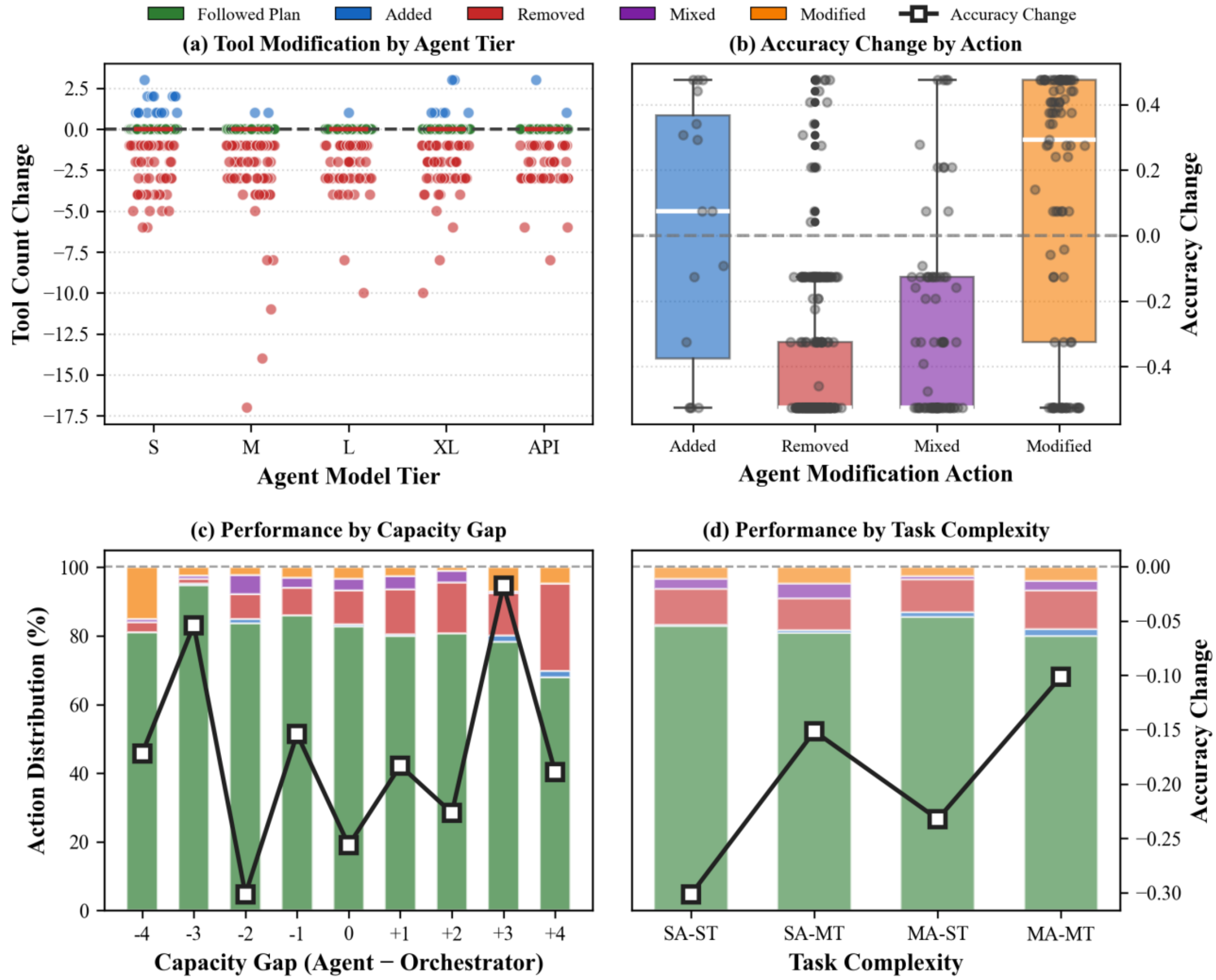

Figure 10. Impact of Orchestrator-Specialist Agent Coordination on Multi-Agent Benchmark Capability.

To understand the dynamics between planned and executed tool usage, this section examines how specialist agents interpret and execute orchestrator-generated plans across 2,650 centralized test cases.

- **High plan adherence:** Overall, specialist agents followed orchestrator plans in 82.7% of test cases. The L tier showed the highest plan adherence (87.9%), while the smallest models (S tier) showed the lowest (77.9%). The remaining 17.3% exhibited plan deviations, revealing when and why agents modify orchestrator instructions. Among deviation cases, tool removal was the dominant behavior (59.0% of deviations), followed by parameter modification (20.3%), mixed modifications (17.2%), and tool addition (3.5%). This distribution suggests that specialist agents more frequently perceive orchestrator plans as over-specified rather than under-specified.
- **Local reasoning generally hurts accuracy:** Plan deviations produced consistently negative effects on task accuracy. Cases where agents followed the orchestrator plan achieved a mean accuracy of 52.5%, whereas deviated executions dropped by 26.2% ($t = -11.335$, $p < 0.0001$). Among deviation types, tool removal resulted in the largest accuracy degradation (−39.1%, $p < 0.001$), followed by mixed modifications (−30.3%, $p < 0.001$). Interestingly, parameter modification was the only deviation type that improved accuracy (+9.9%, $p = 0.045$), suggesting that agents can effectively fine-tune execution details when the overall plan structure is preserved. Tool addition showed no significant effect (+1.9%, $p = 0.87$).

- **Capacity gap effects:** The relative capability difference between orchestrator and specialist agents (specialist tier minus orchestrator tier) significantly influenced plan adherence ($p = 0.024$). When specialist agents were smaller than the orchestrator, plan adherence reached 86.6%. In contrast, when specialists were larger than the orchestrator, adherence decreased by 8%, indicating that more capable specialists are more likely to override orchestrator decisions. However, such overrides consistently degraded accuracy across all capacity-gap levels, with accuracy drops ranging from −2.2% to −37.5% when agents deviated. Notably, the most severe penalties occurred at moderate negative gaps (gap = −2), where smaller agents with limited intelligence attempting to modify plans from stronger orchestrators produced particularly poor outcomes.
- **Task Complexity Independence**. Task complexity had limited influence on plan-adherence rates. Single-agent tasks (SA–ST, SA–MT) showed 82.6% adherence, similar to multi-agent tasks (MA–ST, MA–MT) at 83.1%. However, the accuracy drop associated with deviation varied with task complexity: simpler tasks (SA–ST) experienced larger accuracy drops (−30.1%) than the most complex tasks (MA–MT, −10.1%).

## 4. Discussion

### 4.1 Lessons Learned from Benchmarking Agentic AI Systems

#### 4.1.1 Lesson 1: Larger Models are not always More Reliable

Although larger models are generally stronger in reasoning and planning, our benchmark reveals a critical deployment bottleneck: extra-large models can be more fragile under constrained or volatile compute resources. In practice, infrastructure-driven failures (e.g., latency spikes, timeouts, resource contention) may dominate workflow reliability instead of “bigger is better” trend.

This motivates a systems-level requirement: scalable agentic AI must be compute-aware, rather than purely capability-driven. Effective strategies include adaptive timeouts, retries, request scheduling, caching, and graceful degradation (e.g., falling back to smaller models when contention is detected). More broadly, this lesson suggests that in realistic deployments, the “best” model tier is not necessarily the largest, but the one that consistently completes workflows under operational constraints.

#### 4.1.2 Lesson 2: Architecture Design Matters More than Prompts

Our benchmark indicates that architecture design can influence performance more strongly than prompt optimization. Intelligence modes (centralized vs. decentralized, one-stage vs. two-stage planning) directly shape workflow accuracy and efficiency because they determine how much cognitive load is concentrated in the orchestrator versus distributed across specialist agents.

A key scalability challenge is prompt explosion. As the number of tools and agents grows, providing complete schemas and descriptions in a single planning prompt becomes expensive, noisy, and error-prone. Two-stage planning addresses this issue by compressing the decision space: the orchestrator first performs lightweight routing, and then retrieves schemas only for the selected tools to generate executable, parameterized instructions.

Looking forward, future tool ecosystems may involve **hundreds of agents and thousands of tools**, where even two-stage planning may face complexity limits. This motivates a more scalable direction: hierarchical organization inspired by human companies. Agents can be grouped into departments with clear responsibilities and tool boundaries, with department-level managers acting as intermediate routers. This hierarchical decomposition reduces global complexity and makes coordination more modular. Notably, this direction is aligned with emerging industrial trends such as skill-based tool ecosystems (e.g., *Claude skills*), which organize capabilities into modular “skills” and emphasize routing and retrieval to avoid exposing the full tool space to a single planner.

#### 4.1.3 Lesson 3: Misalignment Between Orchestrator and Specialists Can Increase Cost and Hurt Correctness

Our benchmark shows that the balance between the orchestrator and specialist agents is critical to end-to-end performance. In many runs, a strong orchestrator already produced a correct and well-structured execution plan. In such cases, scaling specialist models does not necessarily improve outcomes.

Lightweight specialists perform well by following the plan, while strong specialists may introduce unnecessary overhead and increase inference cost without clear accuracy gains. More importantly, we observe a failure mode where mid-tier specialists attempt to “re-reason” the orchestrator’s plan, but lack sufficient intelligence to do so correctly. This can lead to deviations from the intended workflow, incorrect tool usage, or broken dependencies, ultimately reducing accuracy despite higher cost. These results suggest that agentic system configuration should not follow the naive rule of “using the strongest model everywhere.” Instead, model pairing should be treated as a role-based allocation problem: the orchestrator handles workflow-critical decisions (task decomposition, routing, dependency management), while specialists focus on localized execution (parameter extraction and tool calling). Therefore, a practical guideline is: invest in orchestration first, and scale specialist capability only when needed.

#### 4.1.4 Lesson 4: Tool Design Should be Agent-Native

A major lesson from execution traces is that tool design itself becomes a first-order factor. Our MCP tools were largely developed based on human engineering habits: flexible nested dictionaries, complex configuration objects, optional fields, and implicit assumptions that are natural for developers. However, for LLM agents, these interfaces can be difficult to use correctly. Deep nesting and complex schemas frequently lead to missing keys, incorrect structure, invalid values, or inconsistent formatting, causing tool execution failures even when the agent’s intent is correct.

This creates an important design gap: tools that are **“human-friendly” are not necessarily “agent-friendly.”** In future agentic platforms, MCP tools should be treated not merely as APIs, but as communication protocols between cognition and execution. Our experience suggests several agent-native design principles, including schema simplification, flattening deeply nested structures, decomposing complex tools into atomic tools, enforcing stronger typing/enumerations, providing validated defaults, and improving tool documentation specifically for LLM interpretability.

More broadly, this motivates a bidirectional workflow for tool development: rather than only asking humans what tools agents should use, we should also ask agents what tools they want. Tool failure logs and agent feedback can guide iterative redesign toward agent-native abstractions, enabling scalable tool ecosystems with lower error rates.

#### 4.1.5 Lesson 5: Human–Agent Alignment Requires Co-Adaptation

Beyond model capability and architecture, our benchmark suggests that agentic AI reliability is also strongly shaped by how users express tasks. Unlike traditional software, agentic workflows are sensitive to problem framing: ambiguous, underspecified, or multi-objective requests can induce early-stage planning and routing errors that propagate through downstream tool execution. This effect becomes more pronounced under constrained compute settings and lightweight local models, where reasoning and recovery capacity is limited.

This motivates an important perspective: alignment is not only an agent-side problem, but also a **human–agent co-adaptation process**, such as knowledge schema alignment, autonomy and agency alignment, operational alignment and training, reputational heuristics alignment, ethics alignment and human engagement alignment [36]. Users gradually learn how to provide clearer goals, constraints, and expected outputs, while agents must learn to actively request missing information and structure tasks into executable steps. Therefore, improving real-world usability requires not only better models, but also better interaction design[37], such as structured request templates, clarification loops, and progressive guidance so that human intent can be reliably translated into tool-driven workflows.

#### 4.1.6 Lesson 6: Toward Dynamic and Self-Evolving Agentic AI Systems

A key lesson from this study is that most existing agentic frameworks rely on predefined workflows (e.g., LangGraph [38]) which limits adaptability in real building applications , where user intents, tool registries, and operating conditions evolve continuously. More dynamic frameworks such as AutoGen [39], CrewAI [40] and this work support planner-driven orchestration and flexible task decomposition, however, they are still constrained by fixed agent pools, predefined role schemas and agentic structure. Therefore, such static or semi-static agentic workflows are insufficient for long-term autonomous operation.

In this work, we take an initial step toward more dynamic agentic systems via an "HR-like" mechanism: when a task cannot be completed with the existing agent pool, the orchestrator can *hire* a new specialist by generating a compliant agent card, or *revise* an existing card to expand its responsibility and tool scope. This reduces dependence on fixed, human-designed agent decomposition and improves robustness under changing tool availability.

More broadly, dynamic agentic AI should be viewed as system-level evolution, including (1) coordination and architecture adaptation where agent revise their structure accordingly, (2) prompt/policy refinement guided by failure patterns, (3) tool co-evolution through schema/interface redesign, and (4) adaptive recovery strategies (retry and fallback toolchains). Overall, agentic AI systems should be treated as evolving organizations, capable of reorganizing roles and procedures as tool ecosystems and task distributions shift over time.

### 4.2 From Rule-Based Workflows to Agentic AI: Motivation and Advantages

Table 6 summarizes the key advantages of agentic AI over classical rule-based automation pipelines. We use the same representative case study from Section 2.3.1 as an example: "How do energy use, operating cost, thermal comfort, and flexibility metrics change if the HVAC COP is upgraded to 4.5 and the battery capacity is increased to 20 kWh for the reference test building? Furthermore, under this upgraded configuration, what is the additional impact of applying a pre-cooling strategy that lowers the zone setpoint by 2°C for 2 hours?"

Table 6 Capability comparison between rule-based automation and agentic AI

| **Factor** | **Rule-based** | **Agentic AI** |
|---|---|---|
| Data Handling | Structured | Fuzzy / Unstructured |
| Decision Logic | IF-THEN, Instruction based | Context-based reasoning |
| Adaptability | Static | Dynamic |
| Transparence | Low | High |

- **Data handling.** Rule-based pipelines typically require strictly structured inputs, meaning users must understand the tool schema (e.g., required parameters, types, and units) and provide complete function-ready inputs. As a result, success becomes highly sensitive to correct parameter extraction and formatting. In contrast, agentic AI is more robust to fuzzy inputs: it can interpret natural language requests, identify the required tool inputs, and automatically extract or infer missing parameters. This significantly lowers the barrier to deploying expert systems to non-expert users, enabling broader accessibility without requiring domain expertise or tool-specific knowledge.

- **Decision logic.** Rule-based approaches execute tasks by following predefined IF–THEN logic. While reliable under well-defined conditions, such systems cannot generalize beyond the instruction space that developers explicitly encode. Agentic AI, instead, can autonomously generate an execution plan, decompose the request into sub-tasks, and route them to appropriate specialist agents or tools. This planning capability is critical for scalability: as user requests become more complex and diverse, it becomes impractical to enumerate all possible workflows using deterministic rules. For example, users may request repeated and nested updates across HVAC systems, DER configurations, and control logic, forming an essentially unbounded instruction space. Agentic AI naturally supports such open-ended interactions through end-to-end reasoning and planning.
- **Adaptability.** Rule-based systems are static by design: they perform well for repetitive and predictable tasks but are fragile when the task or environment deviates from their assumptions. In our case study, the MCP comparison tool only supports comparisons between two cases, whereas the user query requires evaluation across three scenarios (baseline, system upgrade, and control upgrade). A traditional rule-based pipeline cannot resolve this mismatch unless a dedicated workflow is manually added. Agentic AI, however, can adaptively restructure the workflow (e.g., conducting step-by-step comparisons across multiple runs), enabling successful execution even when tool limitations or user intents evolve.
- **Transparency.** Finally, rule-based systems are often opaque to end-users: inputs are processed through predefined scripts, and only final outputs are returned, making it difficult for users to understand intermediate decisions or debug failures. Agentic AI provides higher transparency because it can explain tool usage, reasoning steps, and intermediate results in natural language. This improves interpretability and supports human-centered objectives such as education, monitoring, debugging, and trust building in deployed automation systems.

## 5. Conclusion

This study presented an agentic AI-enabled physics-informed machine learning framework that couples a multi-agent decision layer with the BESTOpt physics-informed runtime environment through Model Context Protocol tool calling. The proposed system supports end-to-end automation of building energy modeling, control, and optimization across multiple domains, including buildings, heating and cooling systems, distributed energy resources, and grid interactions. It translates natural language requests into executable multi-step workflows, coordinates domain specialists, executes simulations, and synthesizes performance impacts across energy use, operating cost, thermal comfort, and flexibility metrics.

A large-scale benchmark consisting of 3,975 cases systematically evaluated the framework under three intelligence structures, 25 large language model configurations, and four task complexity levels, measuring workflow accuracy, token usage, execution time, and inference cost. The results show that centralized two-stage coordination achieves the best accuracy–efficiency tradeoff, because separating planning from execution mitigates prompt explosion and reduces early-stage routing errors. Across model configurations, orchestrator capability dominates end-to-end success. In addition, mid-sized models tend to over-reason execution steps with limited reliability, which can increase deviation from intended plans and degrade overall performance. Finally, orchestrator–specialist misalignment emerges as a major failure source, highlighting the importance of coordination design in scalable agentic workflows.

Looking forward, several directions are critical to advance from research prototypes to deployable autonomous building ecosystems. First, agentic systems should become dynamic and self-evolving, continuously adapting architecture, prompts, and policies based on closed-loop feedback from execution failures. Second, human–agent co-evolution is needed to shift from human-centric tool interfaces toward agent-native workflows, improving tool usability, robustness, and interoperability. Third, scalable hierarchical planning should be further developed to mitigate prompt explosion as tool registries and agent pools grow rapidly. Overall, this study outlines a two-pillar vision that couples agentic AI with a physics-informed machine learning environment for future autonomous, decarbonized, and grid-interactive building energy systems.

## Appendix I: MCP Tools

### I-1: Tool List

MCP tools are grouped into 11 categories as shown in Figure A1:

```
tool_categories:
  - Configuration Related Tools: [config_create, config_save, config_validate, config_set_active,
                                  config_list, config_query, ...]
  - Cluster Related Tools:       [cluster_add, cluster_update, cluster_remove, cluster_query,
                                  cluster_select, ...]
  - Building Related Tools:      [building_add, building_update, building_remove, building_query,
                                  building_select, building_add_thermal_zone,
                                  building_add_electrical_zone, building_add_water_zone, ...]
  - HVAC System Related Tools:   [hvac_add, hvac_update, hvac_remove, hvac_query,
                                  hvac_assign_to_buildings, hvac_select, ...]
  - DER System Related Tools:    [der_add, der_update, der_remove, der_query, der_assign_to_buildings,
                                  der_select, ...]
  - Controller Related Tools:    [controller_add_hvac, controller_add_der, controller_update,
                                  controller_remove, controller_query, controller_assign_to_system, ...]
  - Disturbance Related Tools:   [disturbance_add_weather, disturbance_add_occupancy,
                                  disturbance_add_price,disturbance_update, disturbance_remove,
                                  disturbance_query, disturbance_select, ...]
  - Environment Related Tools:   [environment_add, environment_update, environment_select, ...]
  - Simulation Related Tools:    [simulation_run, simulation_save, simulation_get_status,
                                  simulation_list_results, ...]
  - Analysis Related Tools:      [analysis_comfort, analysis_energy, analysis_cost, analysis_flexibility,
                                 analysis_comprehensive, ...]
  - Comparison Related Tools:    [comparison_comfort, comparison_energy, comparison_cost,
                                  comparison_flexibility, comparison_comprehensive, ...]
```

Figure A1. MCP Tool List

### I-2: An Example MCP Tool Specification

To present how domain functions are packaged as MCP tools, we provide an example specification for adding a new HVAC system to a building cluster as shown in Figure A2. The tool, **'hvac_add'**, is exposed to agents through the MCP server using the **@mcp.tool** decorator, enabling tool discovery and invocation under a standardized MCP interface.

The function accepts two required arguments: **system_id** and **cluster_id**, and three optional arguments: **system_name**, **system_config**, and **parameters**. Each argument is declared with Annotated types and documented via Field (description=...). This design allows the MCP client to export a machine-readable tool schema to the LLM, including argument types and semantic constraints. As a result, the agent can reliably perform tool selection, argument construction, and input validation (e.g., identifier formats, enumerated categories, and units).

The system_config input specifies component-level configurations in a nested dictionary format, covering fan, fan control logic, coil, pump, chiller, and cooling tower. Numeric parameters follow explicit engineering units ($m^3/s$, W, °C), while categorical options such as fan control strategy are constrained to valid enums (i.e., {constant, staged, vfd}). Finally, the tool returns a standardized JSON-like response containing success, data, and message fields (or success and error on failure), facilitating robust integration into multi-agent workflows and runtime logging.

```python
@mcp.tool(description="Add a new HVAC system to a building cluster")
async def hvac_add(
        system_id: Annotated[str, Field(description="Unique identifier for the HVAC system")],
        cluster_id: Annotated[str, Field(description="ID of the cluster to add the HVAC system to")],
        system_name: Annotated[Optional[str], Field(
            description="Display name for the system (e.g., 'FCU System', 'Office HVAC')"
        )] = None,
        system_config: Annotated[Optional[Dict], Field(
            description="HVAC system configuration: {"
                        "'fan': {'rated_flow_m3s': <float m3/s>, 'rated_power_W': <float watts>}, "
                        "'fan_ctrl': {'ctrl_type': 'constant'|'staged'|'vfd', 'rated_flow_m3s': <float>,
                                      'stages': <int for staged mode>}, "
                        "'coil': {'effectiveness': <float 0-1>}, "
                        "'pump': {'rated_flow_m3s': <float m3/s>, 'rated_power_W': <float watts>}, "
                        "'chiller': {'rated_capacity_W': <float watts>, 'rated_cop': <float COP>}, "
                        "'tower': {'rated_capacity_W': <float watts>,
                                   'rated_fan_power_W': <float watts>, "
                        "'pump_power_per_flow': <float W/(m3/s)>, 'min_approach_C': <float °C>,
                         'max_approach_C': <float °C>}}"
        )] = None,
        parameters: Annotated[Optional[Dict], Field(
            description="Additional HVAC parameters beyond system_config"
        )] = None,
) -> Dict[str, Any]:
    """
    Adds a new HVAC system to the specified cluster.

    Example system_config for a typical residential FCU:
    {
        "fan": {"rated_flow_m3s": 0.4, "rated_power_W": 400},
        "fan_ctrl": {"ctrl_type": "constant", "rated_flow_m3s": 0.4},
        "coil": {"effectiveness": 0.7},
        "pump": {"rated_flow_m3s": 0.01, "rated_power_W": 1500},
        "chiller": {"rated_capacity_W": 15000, "rated_cop": 4.5},
        "tower": {
            "rated_capacity_W": 15000,
            "rated_fan_power_W": 400,
            "pump_power_per_flow": 85000,
            "min_approach_C": 3.0,
            "max_approach_C": 7.0
        }
    }

    Fan control types:
    - "constant": Fixed speed fan
    - "staged": Multi-stage fan (requires 'stages' parameter)
    - "vfd": Variable frequency drive for continuous speed control
    """

    # Main functions

        return {
            "success": True,
            "data": {
                "system_id": system_id,
                "cluster_id": cluster_id,
                "system_type": "hvac_systems",
                "system_name": system_name,
                "system_config": system_config,
            },
            "message": f"HVAC system '{system_id}' added to cluster '{cluster_id}'",
        }
    except Exception as e:
        return {"success": False, "error": str(e)}
```

Figure A2. An Example MCP Tool Specification

**Appendix II: Specialist Agents**

**II-1: Agent List**

BESTOpt deploys 11 specialized LLM-based agents as shown below, each defined through structured YAML agent cards specifying role, capability scope, and authorized MCP tool interfaces:

- cluster_agent, building_agent, der_agent, hvac_agent, controller_agent, disturbance_agent, environment_agent, config_agent, simulation_agent, analysis_agent, comparison_agent

**II-2: Agent Card**

Figure A3 presents a representative agent card in YAML format with inline annotations. The agent card serves as a lightweight configuration file used by the agent factory to instantiate an LLM-based agent with a well-defined identity (**agent_id**, **name**), functional scope (**role**, **description**), and generation settings (**model**, **temperature**). The card further specifies the agent's competency boundary via **capabilities**, enforces tool-level permission control through an explicit whitelist (**available_tools**), and provides **example_tasks** as reproducible usage cases. Finally, optional **constraints** encode domain validation rules to improve robustness and prevent unrealistic configurations during tool execution.

```yaml
agent_id:            # Unique identifier for the agent
  hvac_agent
name:                # Human-readable agent name
  HVAC System Configuration Agent
role:                # Agent's functional role in the framework
  Specialist agent for HVAC system configuration and performance parameterization
description:         # Short description of expertise and scope
  Configures HVAC systems, manages HVAC components

model:               # LLM model name used to instantiate this agent
  gpt-4
temperature:         # Sampling temperature for response generation
  0.3

capabilities:        # List of agent capabilities (what the agent can do)
  - Configure HVAC systems and associated components
  - Manage HVAC units and key operational parameters
  - Manage efficiency parameters (e.g., COP) and system-level attributes

available_tools:     # MCP tool IDs this agent is allowed to call
  - hvac_add
  - hvac_update
  - hvac_remove
  - hvac_query
  - hvac_assign_to_buildings
  - hvac_select

example_tasks:       # Example user tasks this agent can solve
  - Add a FCU system for an office building
  - Set chiller cooling capacity to 50 kW
  - Query HVAC efficiency parameters and configuration status
  - Upgrade to a VFD fan

constraints:         # Optional execution constraints
  - Ensure HVAC capacity is consistent with estimated building thermal loads
  - Identify correct system ID before execution
```

Figure A3. Annotated YAML agent card example for the HVAC system configuration agent

**Appendix III: Concierge and Orchestrator**

**III-1: Concierge Agent**

As shown in Figure A4, Concierge is an user-facing interaction agent that (1) clarifies ambiguous requests, (2) normalizes user intents into structured task descriptions, and (3) forwards the structured request to the orchestrator.

```
    SYSTEM_PROMPT = """You are a friendly assistant for a Building Energy and DER (Distributed Energy
Resources) simulation system.

    IMPORTANT: You are a ROUTING layer. The orchestrator behind you has access to default configurations
and is smart enough to handle incomplete requests.

    Your jobs:
    1. Handle greetings and small talk naturally
    2. Route ANY technical/simulation request to the orchestrator
    3. Ask for clarification based on error message if the process failed

    When routing, respond with a brief acknowledgment + ACTION: ROUTE_TO_ORCHESTRATOR

    Examples:
    User: "analyze flexibility after upgrading battery to 20kwh"
    You: "I'll run that flexibility analysis for you. ACTION: ROUTE_TO_ORCHESTRATOR"

    User: "compare energy performance after upgrading HVAC systems"
    You: "Let me compare those for you. ACTION: ROUTE_TO_ORCHESTRATOR"

    DO NOT route:
    - "Hi" / "Hello" / greetings
    - "What can you do?" / capability questions
    - "Thanks" / acknowledgments

    Available agents: {agent_info}
    """

    RESPONSE_FORMATTER_PROMPT = """Format this simulation result for the user in a friendly way.

    Result from system:
    {result}

    User's request: {request}

    Guidelines:
    - Lead with the key finding/answer
    - Show important numbers clearly
    - Keep it concise (5 sentences max)
    - Suggest ONE relevant follow-up if appropriate

    If there's an error, briefly explain and suggest trying again."""
```

Figure A4. Prompt of Concierge Agent

**III-2: Orchestrator Agent with Centralized Intelligence**

The Orchestrator with centralized intelligence mode produces a stepwise plan with tool-level instructions and parameter fields.

- One-stage planning: Provide full agent/tool schemas to produce detailed plans in one pass as shown in Figure A5 (1).
- Two-stage planning:
  -Stage 1 (routing) as shown in Figure A5 (2): Provide only agent identity + tool names; output tools_to_use per step.
  -Stage 2 (parameterization) as shown in Figure A5 (3): Provide schemas for selected tools only; output finalized orchestrator_guidance.

```
Create detailed execution plan for CENTRALIZED mode.

## User Request: {user_request}
## Context: {context}
## Available Agents (with full tool schemas): {agent_info}

Return JSON:
{
  "understanding": "interpretation of user intent",
  "reasoning": "analysis referencing workflow patterns",
  "steps": [
    {
      "step_id": "step_N",
      "agent_id": "target_agent",
      "task": "task description",
      "depends_on": ["prior_step_ids"],
      "orchestrator_guidance": {
        "tool_instructions": [
          {
            "tool": "tool_name",
            "parameters": {"param": "value"},
            "expected_output": "description"
          }
        ],
        "validation": "verification criteria"
      }
    }
  ]
}
```

(1) One-stage planning

```
Create a high-level execution plan. Select which agents and tools to use.

## User Request: {user_request}
## Context: {context}
## Available Agents (with tool names only): {basic_agent_info}

Return JSON:
{
  "understanding": "interpretation of user intent",
  "reasoning": "rationale for agent/tool selection",
  "steps": [
    {
      "step_id": "step_1",
      "agent_id": "target_agent",
      "task": "task description",
      "depends_on": ["prior_step_ids"],
      "tools_to_use": ["tool_name_1", "tool_name_2"]
    }
  ]
}
```

(2) Two-stage planning, stage one

```
Complete the execution plan with specific tool parameters.

## User Request: {user_request}
## Plan from Stage 1: {stage1_steps}
## Tool Parameter Details (for selected tools only): {detailed_tool_info}

For each step, fill in the orchestrator_guidance with exact parameters.

Return JSON:
{
  "steps": [
    {
      "step_id": "step_N",
      "agent_id": "target_agent",
      "task": "task description",
      "depends_on": ["prior_step_ids"],
      "orchestrator_guidance": {
        "tool_instructions": [
          {
            "tool": "tool_name",
            "parameters": {"param": "value"},
            "expected_output": "description"
          }
        ],
        "validation": "verification criteria"
      }
    }
  ]
}
```

(3) Two-stage planning, stage two

Figure A5. Prompt of Orchestrator Agent with Centralized Intelligence

**III-3: Orchestrator Agent with Decentralized Intelligence**

The Orchestrator with decentralized intelligence mode produces a dependency-aware, high-level plan without tool-level details.

```
Create high-level plan for DECENTRALIZED mode.
Agents will decide their own tool usage.

## User Request: {user_request}
## Context: {context}
## Available Agents: {agent_info}

Return JSON:
{
  "understanding": "interpretation of user intent",
  "reasoning": "rationale for agent sequencing",
  "steps": [
    {
      "step_id": "step_N",
      "agent_id": "target_agent",
      "task": "high-level task description",
      "depends_on": ["prior_step_ids"],
      "expected_outcome": "success criteria"
    }
  ]
}
```

Figure A6. Prompt of Orchestrator Agent with Decentralized Intelligence

**III-4: Specialist Agent for Centralized and Decentralized Intelligence**

As shown in Figure A7, in centralized mode, the orchestrator provides detailed tool instructions. The specialist agent performs validation, execution, and synthesis through three phases. In decentralized mode, the specialist agent receives only high-level task descriptions and autonomously determines tool usage. The agent performs planning, execution, and synthesis.

```
You are {agent_name}, a specialist agent.
Role: {agent_role}

The orchestrator has provided these tool instructions for the task:
Task: {task}
Instructions: {tool_instructions}

Your available tools: {available_tools}

Analyze and respond with JSON:
{
  "validation": "valid" or "needs_adjustment",
  "reasoning": "your analysis of the instructions",
  "refined_instructions": [
    {
      "tool": "tool_name",
      "parameters": {},
      "reason": "justification"
    }
  ]
}

If instructions are valid, return them as-is in refined_instructions.
If adjustments needed (wrong params, missing tools, etc.), provide corrections.
```

(1) Specialist agent prompt for centralized intelligence

```
You are {agent_name}, a specialist agent.
Role: {agent_role}

Available Tools: {tool_descriptions_with_schemas}

Task Request: {request}

Analyze the task and plan which tools to use.
Return JSON:
{
  "reasoning": "step-by-step analysis of the task and why you chose these tools",
  "tool_calls": [
    {
      "tool": "tool_name",
      "parameters": {},
      "reasoning": "why this tool with these parameters"
    }
  ]
}
```

(2) Specialist agent prompt for decentralized intelligence

Figure A7. Prompt of Specialist Agent for Centralized and Decentralized Intelligence

**III-5: Dynamic Agent Generation and Adaptation**

As shown in Figure A8, the Orchestrator evaluates whether the current agent pool can handle the user request before planning under dynamic mode. If gaps are identified, the system can create new agents or revise existing agent capabilities.

```
Analyze if current agents can handle this task, or if modifications are needed.

User Request: {user_request}
Context: {context}

Current Agents: {agent_capabilities}
All Server Tools: {all_server_tools}
Tools Not Assigned to Any Agent: {unused_tools}

Return JSON:
{
  "can_handle": true/false,
  "analysis": "explanation of capability assessment",
  "modifications": [
    {
      "action": "create" or "revise",
      "agent_id": "new_or_existing_agent_id",
      "name": "agent name (for create)",
      "role": "agent role (for create)",
      "description": "agent description (for create)",
      "tools": ["tool_names_to_assign"],
      "reason": "justification for modification"
    }
  ]
}
```

Figure A8. Dynamic Mode Prompt for Agent Generation or Adaptation